\documentclass[final,5p,times,twocolumn]{elsarticle}

\usepackage{lineno}

\biboptions{sort&compress}

\usepackage{esvect}
\usepackage{textgreek}
\usepackage{textcomp}
\usepackage{amssymb}
\usepackage[export]{adjustbox}
\usepackage{subcaption}
\usepackage{caption}
\usepackage{wrapfig}
\usepackage{graphicx}
\usepackage{lipsum}
\usepackage{float}
\usepackage{bm}
\usepackage{makecell}
\usepackage{siunitx}
\usepackage{amsmath}
\usepackage{hyperref}

\begin{document}

\begin{frontmatter}

\title{Atomistic survey of grain boundary – dislocation interactions in FCC Nickel}

\author{Devin W. Adams\fnref{label1}}

\author[label1]{David T. Fullwood}
\author[label2]{Robert H. Wagoner}

\cortext[cor1]{Corresponding author}
\author[label1]{Eric R. Homer\corref{cor1}}
\ead{eric.homer@byu.edu}

\address[label1]{Department of Mechanical Engineering, Brigham Young University, Provo, UT, USA}

\address[label2]{Department of Materials Science and Engineering, The Ohio State University, Columbus, OH, USA\fnref{label4}}

\begin{abstract}
It is well known that grain boundaries (GBs) have a strong influence on mechanical properties of polycrystalline materials. Not as well-known is how different GBs interact with dislocations to influence dislocation movement. This work presents a molecular dynamics study of 33 different FCC Ni bicrystals, each subjected to four different strain states to induce incident dislocation-GB interactions for 132 unique configurations. The resulting simulations are analyzed to determine properties of the interaction that affect the likelihood of transmission of the dislocation through the GB in an effort to better inform mesoscale models of dislocation movement within polycrystals. It is found that the ability to predict the slip system of a transmitted dislocation using common geometric criteria is confirmed. Furthermore, machine learning processes are implemented revealing that geometric properties, such as the minimum potential residual Burgers vector (RBV) and the disorientation between the two grains, are stronger indicators of whether or not a dislocation would transmit than other properties, such as the resolved shear stress.
\end{abstract}


\end{frontmatter}



\section{Introduction}
\label{sec1}

It is no secret that mechanical behavior is strongly influenced by the grain boundaries (GBs) within the material. The movement of dislocations, the main carriers of deformation in most metals, is inhibited by the material\textquotesingle s GBs, which can act as obstacles, sinks, or sources for dislocations \cite{Farkas:2013, Kacher:2014, Hasnaoui:2004, Voyiadjis:2016}. This influence can be observed in the Hall-Petch relationship, which shows that as the average grain size of a polycrystalline material decreases its yield strength increases  \cite{Hall:1951, Petch:1953}. 
This relationship has been exploited for years in the production of enhanced materials, such as nanocrystalline metals, which offer significantly increased hardness and strength \cite{Ovidko:2018, Koch:2003, Meyers:2006}. While the Hall-Petch relationship has been shown to match the macroscopic effects of reduced grain size, understanding of exactly how dislocation interactions at each GB cause the Hall-Petch relationship is still not well understood.

The work presented here is part of a collaborative effort involving experimental work \cite{Hansen:2017,Jackson:2016,Ruggles:2013,Ruggles:2016}, mesoscale modeling \cite{Bong:2017,Lim:2011}, and atomistic simulations \cite{Wyman:2017} aimed at better understanding how large populations of dislocations interact with GBs. The present work contributes by investigating the attributes that affect the GB-dislocation interactions at the atomic scale, which can then be used to inform the mesoscale modeling and interpret experimental observations.

Due to the relative paucity of characterized GBs to the vast number of possible GBs, characterizing such interactions remains a daunting task despite the work already done in this field. This stems from the 5-degrees of freedom that define the macroscopic character of a GB: three to define neighboring grains' relative rotation to one another and two to define the boundary plane\textquotesingle s orientation \cite{Olmsted:2009:energy, Sutton:1995}. Additional complexity is involved, since multiple types of GB-dislocation interactions are possible and can be summarized here in four main categories: 1) nucleation of a dislocation at the GB, 2) absorption of a dislocation into the GB, 3) slip transmission wherein the dislocation passes through the GB, and 4) reflection of the dislocation at the GB  \cite{Shen:1988,Sangid:2012,Wang:2015,Wyman:2017,Clark:1989}, with the most studied being nucleation and transmission. Additional attributes, such as temperature, structure of the GB at the location of the interaction, the slip systems involved, etc., further complicate the ability to fully resolve the nature of GB-dislocation interactions. Finally, the time and length scales required to study dislocation-GB interactions make it difficult to capture all the attributes involved in such an interaction in a small number of experiments or simulations. Despite the daunting size of the task, several studies have made significant strides in understanding dislocation-GB interactions, as discussed below.
    
From experimental work, researchers have been able to gain a better understanding of the attributes involved in the dislocation-GB interactions responsible for effects such as the previously mentioned Hall-Petch relationship. Techniques such as transmission electron microscopy (TEM), electron back-scattered diffraction (EBSD), and digital image correlation (DIC) allow one to observe the dislocation activity present within a strained specimen and capture attributes involved in dislocations' interactions with GBs \cite{Shen:1988,Abuzaid:2012,Lim:1985, Koning:2003}. For example, Shen et al. observed dislocation transmission in 304 stainless steel, finding that geometrically well aligned slip systems at the GB are the preferred slip systems for transmission, with the resolved shear stress (RSS) being the deciding factor for transmission if two or more slip systems were equivalently aligned \cite{Shen:1988}. This work confirmed and refined Livingston and Chalmers' geometric criteria for predicting which slip system would be activated in a transmission event \cite{Livingston:1957}. By adding the stipulation that the residual Burgers vector (RBV) should be minimized to the geometric criteria of Livingston and Chalmers and the stress criteria from Shen et al., the  commonly used LRB criteria  was created \cite{Lee:1989}.  Additionally, Abuzaid et al. used DIC in combination with EBSD to support the hypothesis that the RBV is frequently minimized when transmission occurs 
\cite{Abuzaid:2012}. Lim and Raj observed more slip continuity through GBs with low $\Sigma$ value coincident site lattice (CSL) GBs as opposed to those with high $\Sigma$ values \cite{Lim:1985}. While significant understanding has been achieved through experimental techniques, these methods are limited in their capabilities  to control the interactions observed. Two main shortcomings that restrict the abilities of experimental techniques to more fully explore this problem are their inability to observe interactions in a large variety of GBs, as well as the challenge of measuring additional attributes associated with the interaction, such as GB energies or the structure of the GB at the location of the interaction.  

Researchers have developed mesoscale models, like the discrete dislocation dynamics model \cite{Weygand:2002, Zhang:2017, Li:2009} to model the dislocation-GB interactions seen experimentally and have shown that the Hall-Petch relationship is dependent on the ability of dislocations to transmit. Another model, developed by Lim et al., utilizes a two-scale model called the Superdislocation (SD) model, to model the Hall-Petch effect in polycrystals using Finite Element Method (FEM) techniques \cite{Lim:2011}. This method builds on work done by Shen et al. \cite{Shen:1986} to determine a GB\textquotesingle s resistance to dislocation absorption or transmission by calculating an effective critically resolved shear strength of the GB, termed the obstacle stress, \texttau\textsubscript{obs}, for a given dislocation-GB interaction, according to the equation

\begin{equation}
    \tau_{obs} = (1-TF)\tau^*
    \label{eqn:Tau_obs}
\end{equation}

\noindent where $\tau^*$ was observed in stainless steel to be approximately five times the macroscopic yield strength and TF is the transmissivity factor (not to be confused with the Taylor Factor) which measures the relative alignment between the impinging dislocation slip system and the potential transmitted dislocation slip systems. In its current state, the SD model shows promise in predicting deformation behavior when the material response is based on the interactions between the dislocations at the microscale \cite{Lim:2011}. 
The authors believe that the SD model's accuracy could be improved by inclusion of better information about GB-dislocation interactions.

Molecular dynamics (MD) tools provide a complementary approach to exploring the variety of attributes believed to affect the resulting interaction and to evaluate different criteria for transmission of dislocations. In a variety of MD simulations, a number of factors have been found to affect how dislocations interact with GBs, including: the static energy of the GB \cite{Sangid:2011}, ratio of RSS from outgoing dislocations to incoming dislocation \cite{Koning:2003}, Schmid vs. non-Schmid slip \cite{Dewald:2007:Screw}, temperature  \cite{Chandra:2015}, misorientation \cite{Koning:2003, Koning:2002,Swygenhoven:2006, Bachurin:2010}, and location of the interaction \cite{Bachurin:2010}. However, in a separate study, Mrovec found that the geometric criteria commonly used to characterize transmission do not always hold \cite{Mrovec:2009}.

Of particular interest in this work is the ability to predict which slip system the emitted dislocation will transmit onto when transmission occurs. Several MD studies have confirmed experimental findings which suggest that selection should be made based upon the potential slip system with the maximum TF or upon minimization of the magnitude of the RBV left in the GB after transmission\cite{Sangid:2011,Sangid:2012,Koning:2003,Koning:2002}. 

Also of interest is the ability to predict what kind of reactions will occur when an incident dislocation impinges on a GB. In several instances, researchers have suggested equations that help to determine the capability of a dislocation to transmit. For example, Sangid et al. were able to show that, in agreement with experiments \cite{Lee:1989, Lee:1990, Shen:1988}, the energy barrier to transmission is higher for low-energy GBs and lower for high-energy GBs \cite{Sangid:2011}. In a separate study, Li et al. demonstrated that a critical penetration stress for a dislocation can be calculated as a function of the grain boundary energy, the shear modulus, and the RBV \cite{Li:2009}. Although these models are effective in describing the roles that the respective attributes play in transmission, these studies focus more on creating rules to describe the observed interactions rather than creating rules to predict whether such reactions should have occurred. With the range of current results in the study of dislocation-GB interactions, some conflicting and some agreeing, it is apparent that this problem is far from resolved. To complement these studies, the current work is focused on studying a large population of dislocation-GB interactions in order to determine which attributes are important across the whole data set in affecting the interaction.

Much of the previous simulation work has focused on smaller numbers of GBs with single dislocation-GB interactions. The current work seeks to explore these behaviors in a larger study of dislocation-GB interactions utilizing  a subset of Olmsted's 388 Ni bicrystals \cite{Olmsted:2009:energy}. The manuscript first describes the methods to simulate and analyze dislocation-GB interactions in the MD simulations of Ni bicrystals. In the results, a detailed analysis of a simulation is first demonstrated after which the large number of GB-dislocation interactions are examined. This is followed by a discussion of the results and their potential to improve models that describe GB-dislocation interactions. The work concludes with a machine learning model to predict transmission of dislocations through a GB.

\section{Method}
\label{sec2}
In order to better resolve some of the attributes that may influence a dislocation-GB interaction, individual interactions in simplified systems are desired. To simulate such interactions, Ni bicrystals with a flat and well-defined GB plane are loaded using the open source molecular dynamics code produced by Sandia National Laboratories, LAMMPS (Large Scale Atomic/Molecular Massively Parallel Simulator)  \cite{Plimpton:1995}. During the simulation, dislocations periodically emerge from a notch and are driven towards the GB. The resulting analysis of each interaction observed in the simulations enables the examination of both geometric (i.e., the TF, RBV, and misorientation in this study) and non-geometric (i.e., the RSS and the static GB energy) attributes.

\subsection{Bicrystals}
\begin{figure}[t]
\includegraphics[width=.75\linewidth]{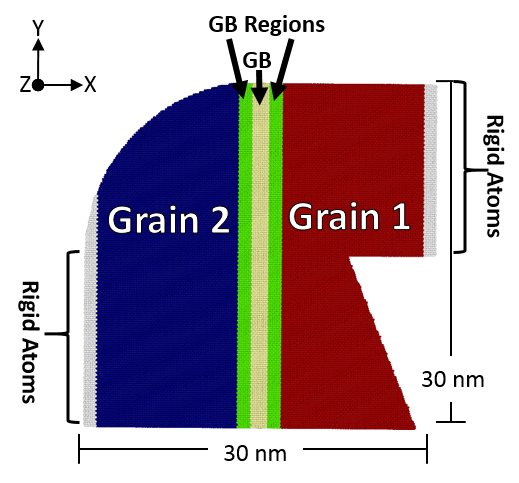}
\centering
\caption{Diagram of a standard bicrystal configuration used in the MD simulations. The two grains form a planar GB in the center of the bicrystal. Regions of atoms 2-6 lattice parameters (7-18\AA) on either side of the GB are shown in light green and are used to calculate the different properties of the dislocation-GB interaction. The average dimension for each bicrystal is 30x30x7nm with a total of approximately 5x10\textsuperscript{5} atoms. The bicrystal is pulled in tension along the X direction by applying a tensile force to the rigid body of atoms on either side of the bicrystal.}
\label{fig:Bicrystal_Diagram}
\end{figure}

A subset of 33 different bicrystals is chosen from the set of 388 minimized Ni bicrystals created by Olmsted et al. \cite{Olmsted:2009:energy}. To create the 388 GBs, Olmsted et al. found all possible GBs that have a periodic boundary interface within a cell size of L\textsubscript{max}=15a\textsubscript{o}/2 where a\textsubscript{o} is the lattice parameter. This resulted in 72 unique misorientations from which the 388 GBs were constructed. The selected bicrystals from this set of GBs are all symmetric tilt or symmetric twist GBs about the [100], [110], or [111] disorientation axes and cover a range of disorientation angles and corresponding static GB energies; a complete list of all 33 bicrystals is available in Supplemental Table S1. To induce slip on a variety of slip systems, each bicrystal is rotated 0\textdegree , 90\textdegree , 180\textdegree , and 270\textdegree ~around the GB plane normal prior to the construction of the simulation cell. Because the original bicrystals created by Olmsted et. al are of insufficient size to study dislocation-GB interactions, they were enlarged by adding atoms along the x-direction according to its periodic length in the x-direction, and then similarly enlarged in the y- and z-directions. The simulation cell is created with a notch in one grain, by simply removing atoms, to act as a stress concentrator, the geometry of which is illustrated in Figure \ref{fig:Bicrystal_Diagram}. Supplemental Figure S1 illustrates the construction process for the simulation cells. The simulation cell geometry also has a rounded edge in the opposite grain to discourage dislocation activity in that grain. This geometry is somewhat similar to that used by de Koning \cite{Koning:2002}. 

With 33 bicrystals each rotated to 4 different orientations there are 132 unique simulation configurations. The average dimensions of each cell are approximately 30x30x7nm, containing approximately 5x10\textsuperscript{5} atoms. 

\subsection{Molecular Dynamics Simulation}
Once the bicrystal geometries shown in Figure \ref{fig:Bicrystal_Diagram} are created, the structure is minimized using the conjugate gradient method and then equilibrated for 175ps to a simulation temperature of 10K using an NVT ensemble where the number of atoms, the volume, and the temperature are held constant. The Foiles-Hoyt EAM potential \cite{Hoyt:2005} is implemented as it is the potential used to create the Olmsted GB set \cite{ Olmsted:2009:energy}. Furthermore, this potential has been used to examine GB-dislocation interactions in a few cases \cite{Sangid:2012, Wyman:2017, Sangid:2011}~and shows good agreement with experimental values of intrinsic and unstable stacking fault energies \cite{Siegel:2005}, the latter of which has been shown to be important in the nucleation and mechanics of dislocations \cite{Rice:1992}. Non-periodic boundaries are implemented in all three dimensions to eliminate any potential bias against nucleation on slip systems with long periodic dimensions. 

After equilibration, a constant tensile force is applied on rigid groups of atoms on either end of the bicrystal inducing a strain state on the bicrystal withan average strain rate of 7x10\textsuperscript{8}s\textsuperscript{-1}. The high strain rate is common in MD simulations, which for this type of study has been shown to give relatively equivalent results for a strain rate in the range of 10\textsuperscript{8} - 10\textsuperscript{10}s\textsuperscript{-1} \cite{Sangid:2011}. The tensile force is applied for up to 250ps, with the observed dislocation-GB interactions typically occurring within the first 150ps. The centrosymmetry parameter, Voronoi volume, slip direction, potential energy, and the Virial stress tensor (averaged over the previous 0.3ps) are output every picosecond for all atoms within five lattice parameters of the GB and for all other atoms with a centroysmmetry value greater than 1.0 which captures defects, such as dislocations, in the system. Since the GB structure influences the stress on the atoms immediately surrounding it and we desired to know the stress on the dislocations, only atoms 2-5 lattice parameters (or 7-18\AA) away from the GB plane are considered in the subsequent calculations; this region is indicated by the green regions of atoms in Figure \ref{fig:Bicrystal_Diagram}. Similarly, atoms within 2 lattice parameters of any free surface are ignored in subsequent calculations to reduce the influence of the free surface.

\subsection{Analysis Techniques}
Because of the potential uniqueness of each interaction, substantial effort is made to create a systematic and objective method to analyze the interactions. We first define the methods used to identify the incident and emitted dislocations. This is followed by a brief description of the various types of events observed and identified at the GB. Finally, we detail the different geometric and non-geometric attributes that are recorded for each interaction.

\subsubsection{Incident/Emitted Dislocations}
While similar studies have focused primarily on full dislocations only \cite{Shen:1988,Abuzaid:2012,  Livingston:1957, Dahlberg:2017}, for the FCC nickel bicrystals we observe slip along \{111\} planes in both $\langle110\rangle$ (full dislocation) and $\langle211\rangle$ (partial dislocation) directions. Therefore, each interaction is identified as involving either a partial or a full dislocation. This decision affects the analysis detailed below so the analysis is run twice, once examining only the full dislocation interactions and again examining both the partial and full dislocation interactions. It is worth noting that in several cases (25) we observe the recombination of leading and trailing partial dislocations in the GB. When this recombination occurs before any dislocation emission at the GB, the incident dislocation is classified as a full dislocation.

\begin{figure}[t]
\begin{center}
   \includegraphics[width=0.75\linewidth]{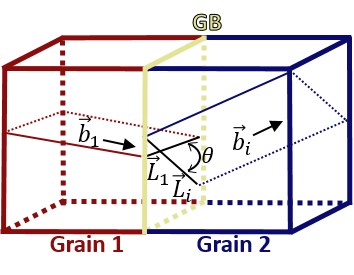} 
\end{center}
\caption{Example transmission event involving an impinging dislocation in grain 1 with a slip direction of $\vec{b}_{1}$ and a line intersection with the GB of $\vec{L}_{1}$. The emitted dislocation travels in grain 2 in the $\vec{b}_{i}$ direction and has a line intersection with the GB of $\vec{L}_{i}$. The angle between $\vec{L}_{1}$ and $\vec{L}_{i}$ is $\theta$.}
  \label{fig:RBV_and_TF_graphic}
\end{figure}

\subsubsection{GB Events}
Simulations and the resulting dislocation-GB interactions are visualized using the OVITO (Open Visualization Tool) software \cite{Alexander:2010} and each timestep of the simulation and its dislocation-GB interactions is observed. These interactions are classified as either transmission, reflection, or absorption, depending on what happens first. Transmission occurs when a dislocation emits from the GB and propagates at least 7\AA~into the body of the second grain without getting reabsorbed back into the GB after nucleating at or near the point of impact from the dislocation in grain 1. Reflection of a dislocation is defined in a similar way to transmission except that following absorption of an incident dislocation, the GB emits a dislocation \textit{back} into grain 1. Finally, a dislocation is classified as having been absorbed when the incident dislocation is absorbed and neither transmission nor reflection are subsequently observed. 
All subsequent activity, such as additional transmission events, following each interaction is not tracked because of the difficulty in correlating it with any incident behavior. 
Furthermore, it should be noted here that what occurs in the GB following any event is not tracked due to the complex nature of and the difficulty in characterizing changes in the GB structure, despite the fact that significant activity can occur. For example, in simple GB structures like the twin boundary, one can observe glide of a dislocation in the GB following absorption \cite{Xu:2016}. The GBs in the present work, while highly symmetric, are sufficiently complex in their atomic structure that tracking dislocation activity through the GB is not pursued.

\subsubsection{Geometric Attributes}
As previously discussed, the most common geometric criteria that have been used to describe the dislocation-GB interaction include the alignment of impinging and potentially emitted slip planes and slip directions as well as the disorientation between the two grains. The geometric attributes considered in this study include the residual Burgers vector (RBV), the transmissivity factor (TF), and disorientation angle. The calculation of the RBV and the TF are briefly discussed here. Figure \ref{fig:RBV_and_TF_graphic} illustrates two slip planes, their respective Burgers vectors, and their line of intersection, which are used to calculate the RBV and the TF for transmission of a dislocation from grain 1 to 2.

Upon transmission through a GB, the total Burgers vector of a dislocation is conserved, with a fragment of it typically being trapped in the GB \cite{Sangid:2012}. This fragment is known as the RBV. According to Shen et al. and others \cite{Sangid:2012,Abuzaid:2012,Koning:2003,Lim:1985,Li:2009, Koning:2002}, the dislocation most likely to transmit is the one which minimizes the magnitude of the RBV. The RBV, reported in this study in units of the lattice parameter $a$, is calculated according to
\begin{equation}
    \vec{b}_{residual} = \vec{b}_{1}-\vec{b}_{i}
    \label{eqn:RBV}
\end{equation}
for the Burgers vectors of the impinging dislocation, $\vec{b}_{1}$ and potential emitted dislocation, $\vec{b}_{i}$, defined in the same reference frame. For each interaction, the RBV is used to predict the slip direction of the outgoing dislocation. 

To predict the full slip system, i.e., slip plane and slip direction, the TF is used. Calculation of a TF for each interaction is defined as
\begin{equation}
    TF=(\vec{L}_{1}\cdotp \vec{L}_{i})*(\vec{b}_{1}\cdotp \vec{b}_{i})
    \label{eqn:Trans_factor}
\end{equation}
where $\vec{L}\textsubscript{1}$ and $\vec{L}\textsubscript{i}$ are the line intersections between the GB and the impinging or outgoing slip plane, respectively, $\vec{b}\textsubscript{1}$ and $\vec{b}\textsubscript{i}$ are the slip directions, and TF is the transmissivity factor \cite{Shen:1988}. Each potential slip system is considered and the emitted slip system that maximizes the value of the TF, ranging from 0 to 1, is the most geometrically aligned with the impinging dislocation. Given a set of potential slip systems, the one with the highest TF is the predicted dislocation slip system to transmit \textit{if} transmission occurs \cite{Shen:1988} and can be seen graphically in Figure \ref{fig:RBV_and_TF_graphic}.

\begin{figure*}[t!]
\centering
\begin{subfigure}[c]{0.49\linewidth}
\includegraphics[width=1\textwidth]{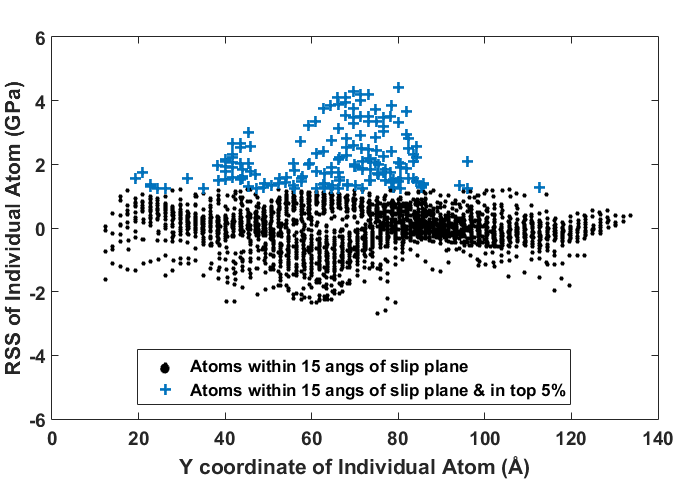}
\caption{}
\label{fig:Atom_stress_distribution}
\end{subfigure}
\begin{subfigure}[c]{0.49\linewidth}
\includegraphics[width=1\textwidth]{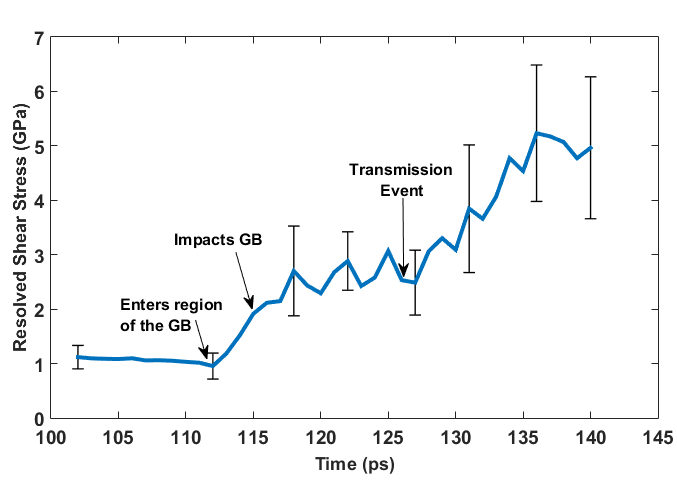}
\caption{}
\label{fig:RSS_error_bars_36p6}
\end{subfigure}
\caption{a.) The RSS on each individual atom that is slipping within the GB region and within 15\AA~of the dislocation normal. The light blue crosses indicate the 5\% of atoms that have the highest RSS. b.) Average RSS, as calculated using the top 5\% of atoms, on a particular dislocation as a function of time. As can be seen, the associated error with the RSS is significant after the dislocation impacts the GB.}
\label{fig:Atom_stress_dist_and_Error_bars}
\end{figure*}

\begin{figure*}[t]
\includegraphics[width=1\textwidth]{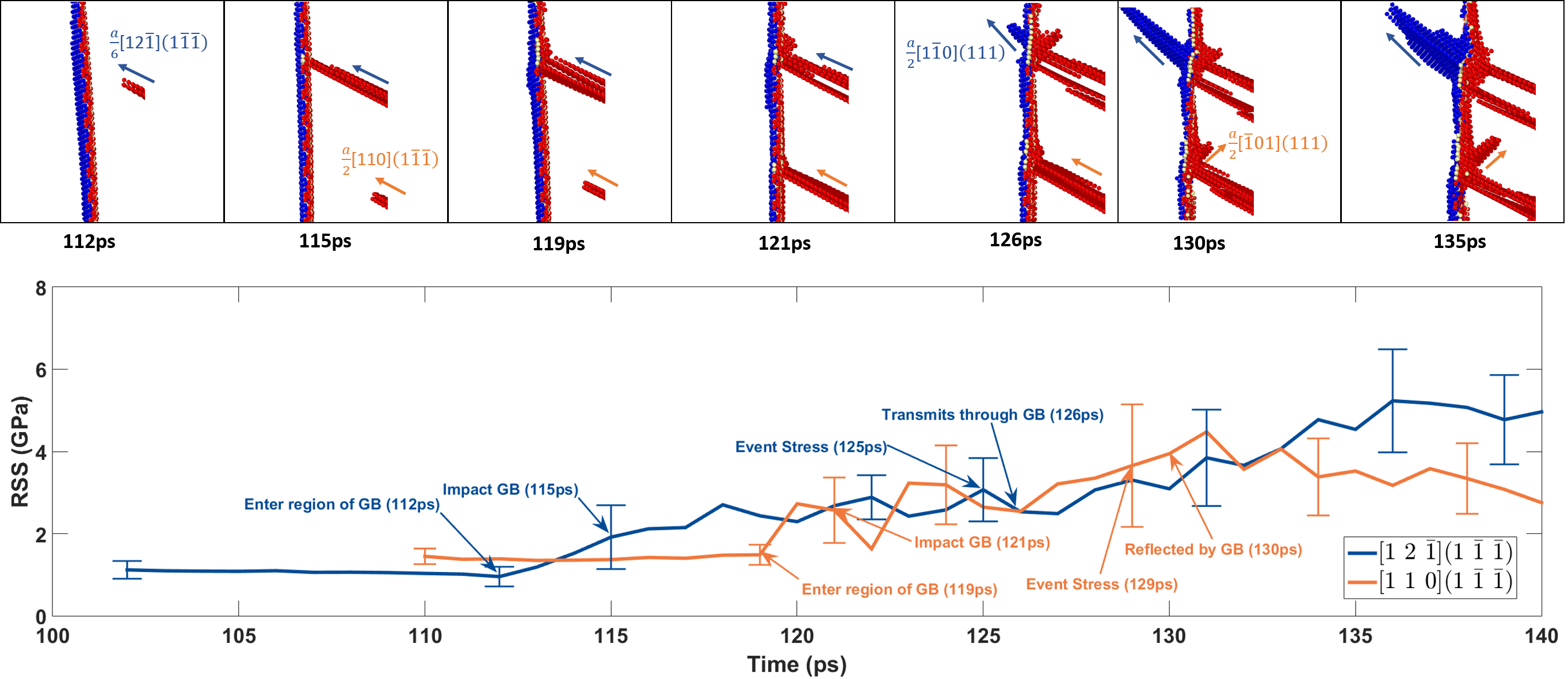}
\centering
\caption{Snapshots of the simulation of the [100] Tilt $\Sigma$25a bicrystal. Two dislocations of the same slip plane but different slip directions impact the GB at different locations and at different times. The first dislocation, indicated by a dark blue label, transmits and the second dislocation, indicated by a light orange label, reflects. Here the RSS associated with each event for the discussed dislocations are shown beneath the snapshots and are labelled accordingly.}
\label{fig:NI36_dislocation_snapshots}
\end{figure*}

The final geometric attribute considered is the disorientation between the two grains. Bachurin et al. and others found that the propensity to transmit dislocations is dependent on the disorientation angle between the two grains,  \cite{Li:2009,Koning:2002, Hamid:2017,  Bachurin:2010,Dahlberg:2017, Aust:1954,  Davis:1966, Gao:2017}. In their study, Li et al. found that this could partially be explained by the fact that the grain boundary energy is a function of the misorientation angle, thus affecting the stress required to push a dislocation through the GB, with increasing misorientation requiring higher stresses \cite{Li:2009}. To test this dependence, GBs with a wide range of disorientation angles are selected for this study.

\subsubsection{Non-geometric Attributes}
Since several publications have shown the static GB energy to correlate with dislocation-GB interactions \cite{Sangid:2011,Sangid:2012} it is examined here as well. The static GB energy is available for each of the GBs as obtained by Olmsted et al. \cite{Olmsted:2009:energy}. 

In addition to the GB energy, we also examine the stresses associated with the dislocation-GB interactions. Specifically, we calculate the resolved shear stress (RSS) on the incident dislocation. From this measured stress we can define an event stress associated with the interaction. The event stress is measured as the maximum RSS that 1) is between the time the dislocation impacts the GB and the time the event occurs \textit{and} 2) occurs within 1-10ps before the event. By imposing these two rules, we are able to maintain a consistent and objective way to determine the event stress. The rationale for picking the maximum stress before the event occurs is because it is believed that the dislocation would be less likely to transmit, reflect, or be absorbed at a lower stress. Therefore, the maximum stress provides an estimate for a potential critically resolved shear stress on a given dislocation required to cause transmission, absorption, or reflection.

Care is taken to calculate the event stress in a manner that minimizes the uncertainty of averaging stress in MD calculations. To demonstrate the uncertainty, Figure \ref{fig:Atom_stress_distribution} shows the stress on individual atoms that are within 15\AA~normal to the slip plane on which a dislocation is traveling. As can be seen, there is a large range of stresses seen in the region of the dislocation, with a bifurcation of the stress visibly present, which occurs near the dislocation core. The two peaks in Figure \ref{fig:Atom_stress_distribution}, positive and negative, show the stress on the dislocation in the region of atoms being measured. The positive stress is used because it represents the atoms that are slipping. Therefore, in order for the bifurcation to not report an average stress around zero, we average the stress of the atoms with the top 5\% RSS values. This average simultaneously reduces the noise present in the analysis and is less sensitive to the stress of nearby dislocations. Even with this filtering process, significant uncertainty is still present, as indicated by the plot in Figure \ref{fig:RSS_error_bars_36p6} where the averaged RSS and its standard deviation of the top 5\% of stress values are plotted as a function of time. This level of uncertainty proves to be a challenge in our efforts to extract correlations of interactions with the event stress. In spite of this, we do find that there may be trends for individual slip systems.

\begin{table}[t!]
\centering
 \caption{Table of all possible TFs and RBVs (in units of lattice parameter, a) for the first interaction shown in Figure \ref{fig:NI36_dislocation_snapshots}, sorted according to the TF. The transmitted slip system is shown in bold font and has the highest TF and lowest RBV for any of the potential full dislocations. For a complete table of all considered full and partial slip systems, see Table S2 in the supplemental material.}
\small
 \begin{tabular}{@{}*{4}{c}@{}} 
 \hline
 \thead{Potential\\Slip System} & \thead{Transmissivity \\Factor, TF} & \thead{Transmitted RBV (a) } & \thead{Reflected RBV (a)} \\ [0.5ex] 
 \hline
$[1\bar{1}2](\bar{1}11)$ & 0.654 & 0.216 & 0.408  \\
$[1\bar{2}1](111)$ & 0.632 & 0.105 & 0.333 \\
$\mathbf{[1\bar{1}0](111)}$ & \textbf{0.596} & \textbf{0.374} & \textbf{0.707}\\
$[01\bar{1}](\bar{1}11)$ & 0.579 & 1.052 & 1.080\\
...&...&...&...\\
$[21\bar{1}](1\bar{1}1)$ & 0.006 & 0.558 & 0.782 \\
$[112](\bar{1}\bar{1}1)$ & 0.004 & 0.479 & 0.624 \\
$[110](1\bar{1}1)$ & 0.003 & 0.827  & 1.080\\
$[011](\bar{1}\bar{1}1)$ & 0.003 & 0.879 & 0.913\\[1ex] 
 \hline
 \end{tabular}
\label{table:dislocation 36.6 RBVs and TFs}
\end{table}

\section{Results}
\label{sec3}
The results are divided into three sections. The first section analyzes two dislocation-GB interactions of a single simulation in detail in order to illustrate the significance and meaning of all the attributes measured during the large number of simulations. Following this, statistics relating to the dislocation-GB interactions are examined. The final section focuses just on transmission events and the attributes involved in transmission.

\subsection{Case Study of Tilt GB}
 To illustrate what individual interactions look like, we examine the 16.26\textdegree~[100] Tilt ($\Sigma$25a CSL with $(0\,\bar{1}\,7)$/$(0\,\bar{1}\,\bar{7})$ boundary plane normals) bicrystal simulation in detail. This simulation exhibits multiple interactions of the same slip plane and contains two of the three types of events analyzed in this study: transmission and reflection. 

 Figure \ref{fig:NI36_dislocation_snapshots} shows snapshots of the simulation at selected times to show the different events associated with the dislocation-GB interactions. Atoms with a centrosymmetry parameter less than 1.0 are not shown for clarity. Figure \ref{fig:NI36_dislocation_snapshots} also labels the slip systems of each dislocation and plots the RSS (as calculated following the procedure outlined in 2.3.4) for each interaction as a function of time.
 
 In the first interaction at 112ps, indicated by a dark blue label in Figure \ref{fig:NI36_dislocation_snapshots}, the activated dislocation is on the $[12\bar{1}](1\bar{1}\bar{1})$ slip system. Here, the first dislocation is followed by an identical partial dislocation (therefore, it is treated as a partial rather than two partials that make a full dislocation) which impacts the GB about 1ps after the first impact. 
 
 The RSS increases as the second partial dislocation approaches and impacts the grain boundary. Between impact and transmission, which occur at 115ps and 126ps respectively, the RSS on the incident slip system reaches a magnitude of approximately 3GPa just before it transmits through the GB onto the $[1\bar{1}0](111)$ slip system in grain 2. This stress of 3GPa is interpreted as the event stress for this dislocation-GB interaction as it is the maximum RSS on the dislocation after impact with the GB \textit{and} within the 10ps before the transmission event.   
 
The TF and the magnitude of the RBV for the 12 potential full and 12 potential partial emitted slip systems are calculated and compared with the actual slip system of the transmitted dislocation. For brevity, Table \ref{table:dislocation 36.6 RBVs and TFs} only lists the values for the slip systems with the 4 highest and 4 lowest TF values; the complete list for all full and partial slip systems is available in Supplemental Table S2. The slip system on which transmission actually occurred is listed in bold in Table \ref{table:dislocation 36.6 RBVs and TFs}. It can be seen that according to the TF and transmitted RBV, there are two more-geometrically favorable partial slip systems that were not activated. However, the transmitted dislocation does have the highest TF and lowest RBV of the available full slip systems. 
\begin{figure*}[t]
\centering
\begin{subfigure}[c]{0.49\linewidth}
\includegraphics[width=1\linewidth, height=6cm]{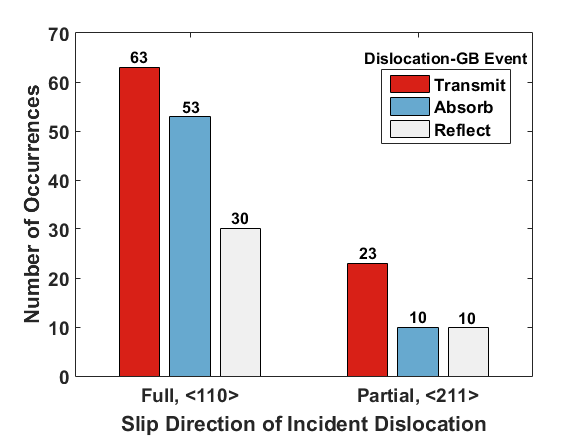}
\caption{}
\label{fig:Classification_incident_dislocations}
\end{subfigure}
\hfill
\begin{subfigure}[c]{0.49\linewidth}
\includegraphics[width=1\linewidth, height=6cm]{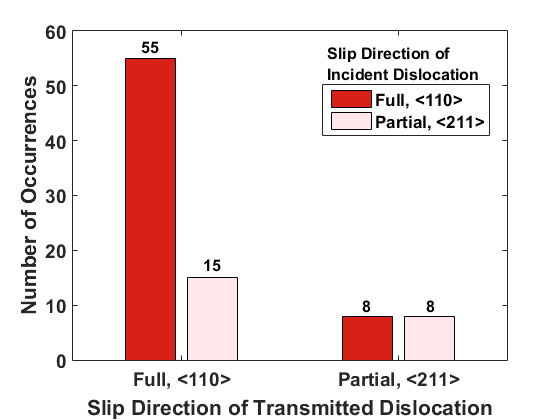} 
\caption{}
\label{fig:Classification_emitted_dislocations}
\end{subfigure}
\caption{a.) Classification of the type of incident dislocations and corresponding dislocation-GB event. As can be seen, the most common event is transmission, followed by absorption with reflection being the least common. b.) Classification of the type of transmitted dislocations and the corresponding type of incident dislocation that transmitted. It is obvious that the vast majority of transmitted dislocations slip along the $\langle110\rangle$ direction and are a result of full dislocations impinging on the GB. }
\label{fig:Classication_of_interactions}
\end{figure*}

During the time of the first interaction, a second dislocation, indicated by a light orange label in Figure \ref{fig:NI36_dislocation_snapshots}, is seen to impact the GB approximately 70\AA~away from the first dislocation and closer to the middle of the bicrystal. This second dislocation impacts the GB at 121ps and involves an incident dislocation of the $[110](1\bar{1}\bar{1})$ slip system.  For this second interaction, the stress in the region continues to rise to $\sim$3.7GPa immediately before the dislocation is reflected back onto the $[\bar{1}01](111)$ slip system in grain 1. Curiously, although it has the same slip plane as the first dislocation, a potential transmission slip system with a higher maximum potential TF of 0.627, a smaller potential RBV of 0.2, and a higher RSS than the first interaction discussed above, this dislocation interaction causes a dislocation to reflect back into grain 1. 
 
 The case study is useful in illustrating the complex nature of these interactions. In contrast to some previous works, we aim to analyze numerous dislocation-GB interactions across a number of different GBs. This enables an informed statistical study to be performed with the trade off of having less control of the interactions present in the simulations and higher uncertainty in measuring attributes like the RSS.

\begin{figure*}[!t]
    \centering
    \includegraphics[width=1\linewidth]{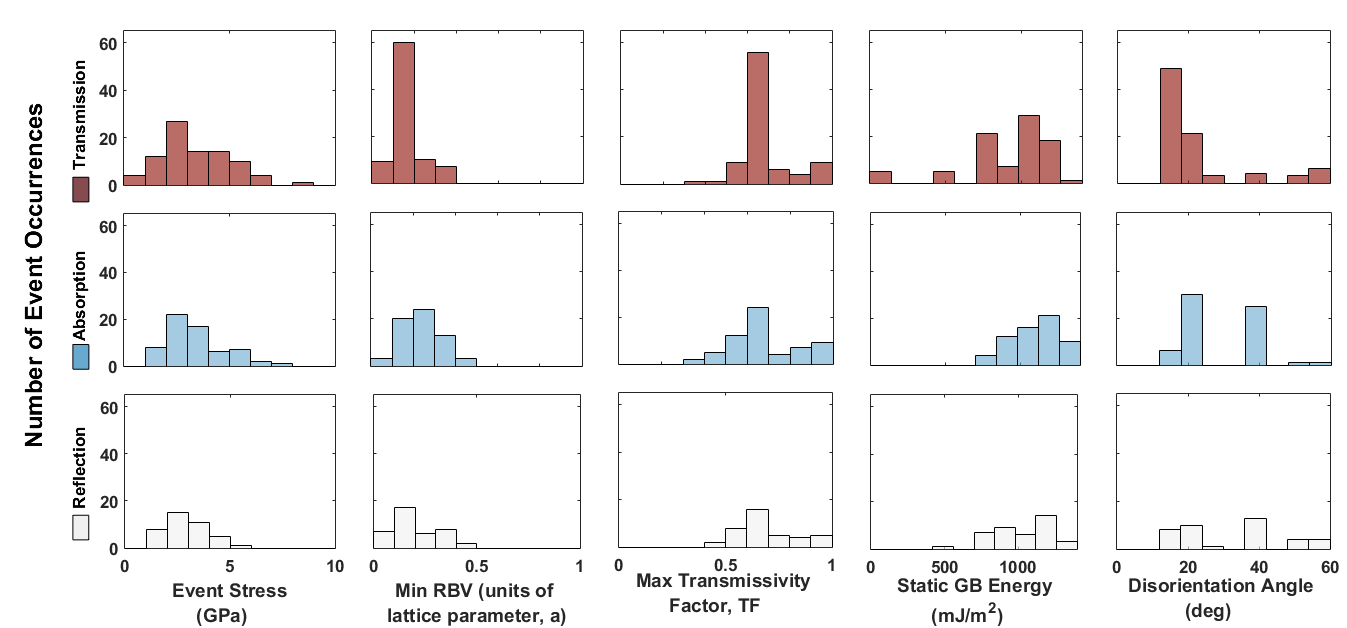}
    \caption{Histograms showing the distribution of the Event Stress, Minimum RBV, Maximum Transmissivity Factor, Static GB Energy, and Disorientation Angle separated for each different event type: Transmission, Absorption, and Reflection. The Minimum RBV and Maximum Transmissivity Factor were calculated considering both full and partial dislocations as potential emitted dislocations.}
    \label{fig:RSS_RBV_TF_GBeng_Diso_histograms}
\end{figure*}

\subsection{Entirety of Interactions}
In all 132 simulations, 189 meaningful dislocation-GB interactions were observed for 31 of the 33 different GB structures tested. The number of dislocation interactions with each unique GB ranges from 2 to 13 with an average of 6 interactions, but since each GB has 4 simulations for different orientations of the notch, the average per simulation is 1.5 interactions. These interactions are classified as either transmission, reflection, or absorption, depending on which happens first and are subsequently analyzed in the same way presented in the case study.

Overall, 86 transmission events, 63 absorption events, and 40 reflection events are observed. These interactions are further sorted, as shown in both Figures \ref{fig:Classification_incident_dislocations} and  \ref{fig:Classification_emitted_dislocations}, into the type of incident dislocation for a given interaction based on its respective slip direction: $\langle110\rangle$ (full) or $\langle211\rangle$ (partial). 

\begin{figure*}[!t]
\centering
\begin{subfigure}[c]{0.49\linewidth}
\includegraphics[width=1\textwidth]{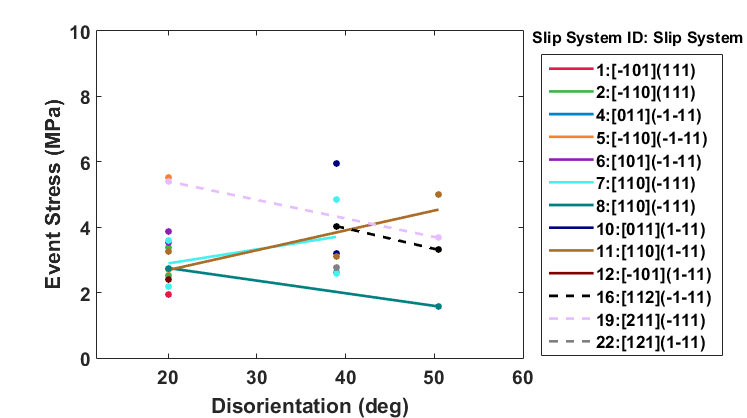}
\caption{}
\label{fig:Event_Stress_vs_dis_vs_SS_110Tilt}
\end{subfigure}
\begin{subfigure}[c]{0.49\linewidth}
\includegraphics[width=1\textwidth]{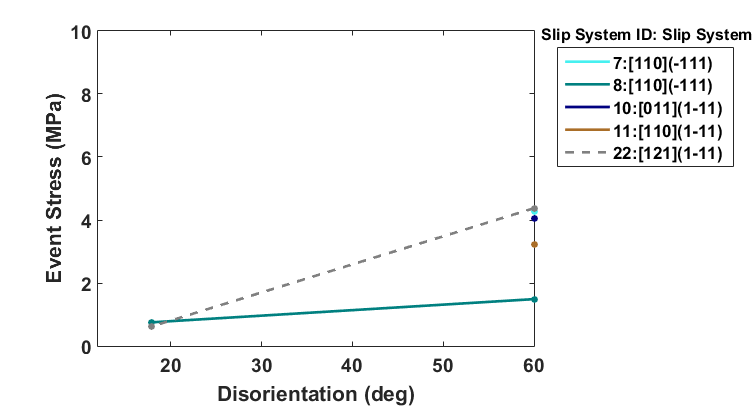}
\caption{}
\label{fig:Event_Stress_vs_dis_vs_SS_111Twist}
\end{subfigure}
\caption{Event stress vs disorientation for activated slip systems for a.) bicrystals with [110] Tilt GBs and b.) biscrystals with [111] Twist GBs. There seems to be some degree of dependence of the event stress on the slip system, as evidenced by the event stress having a positive correlation with the disorientation angle for slip systems 11, 7, and 22 but a negative correlation for slip systems 16 and 19. A more controlled study with more data is required to elucidate such relationships with confidence.}
\label{fig:Event_Stress_vs_dis_vs_SS}
\end{figure*}

 As is shown in Figure~\ref{fig:Classification_incident_dislocations}, the most common event is transmission with absorption a close second. While reflection events are the least common, the large number of events was surprising, not least of all because of the large RBV required for reflection. Examples of the large RBV vectors can be seen in Table \ref{table:dislocation 36.6 RBVs and TFs}. When examined by incident dislocation type, it can be seen that 53.5\% of partial dislocations transmitted and 43.2\% of full dislocations transmitted. The majority of incident and transmitted dislocations were full dislocations, accounting for 77.2\% of all 189 incident dislocations and 81.4\% of all 86 transmitted dislocations.

As discussed in section 2.3, the event stress, minimum RBV, and maximum TF are calculated for each dislocation-GB interaction; both full and partial dislocations are considered in determining the minimum RBV and maximum TF. Histograms of these values, along with  the properties of the GBs involved in the interactions, i.e., GB energy and disorientation angle, are given in  Figure \ref{fig:RSS_RBV_TF_GBeng_Diso_histograms}. The values are divided between the types of events to gauge whether any correlations exist. From these plots, it is easy to see that, while there are significant populations present for each attribute considered, there is too much overlap between the respective values to discriminate between the interaction types based on any single attribute. In other words, even with the large number of events, none of these attributes can be used alone to predict the type of event that will occur.

Since the original motivation of the study was to determine whether a critical GB obstacle stress was associated with transmission events, this aspect is analyzed in more detail. Due to the noise present in the measurements of RSS as a result of the unavoidable interference from nearby dislocations, correlations are difficult to glean. However, potential correlations between the event stress and the event are found when comparing the event stress for activated slip systems as a function of disorientation angle. In Figure \ref{fig:Event_Stress_vs_dis_vs_SS_110Tilt} it can be seen that there is scatter among the data even for events on the same slip system. However, for three slip systems in [110] tilt GBs and two slip systems in [111] twist GBs, in Figures \ref{fig:Event_Stress_vs_dis_vs_SS_110Tilt} and \ref{fig:Event_Stress_vs_dis_vs_SS_111Twist} respectively, we are able to show the existence of some trend with disorientation angle. With the little data we have at present, we are unable to draw definitive conclusions. It is possible that alternate approaches, such as that performed by Wyman et al. to investigate dislocation nucleation from GBs \cite{Wyman:2017}, could elucidate new criteria related to transmission.

\subsection{Transmission}
A significant focus of this research is on the attributes involved in the transmission of dislocations. As such, the remainder of this section emphasizes results found relating to transmission vs no transmission (i.e., absorption and reflection). The first matter addressed is the dependence of the frequency of transmission on different attributes. This is followed by the ability of the TF and the RBV to predict a transmitted dislocation's slip system.

\begin{figure*}[t]
\centering
\includegraphics[width=1\textwidth]{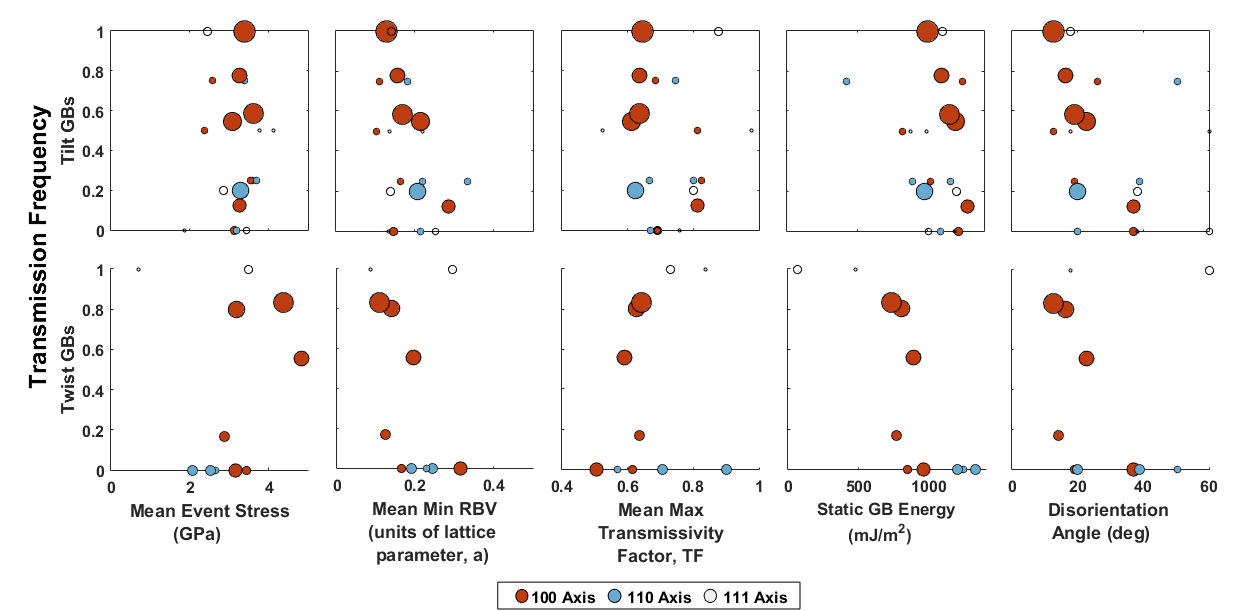}
\caption{Frequency of transmission as a function of mean event stress, mean minimum RBV, mean max TF, static GB energy, and disorientation. The size of each point correlates to the number of data points used to calculate the frequency of transmission for the corresponding x value.  There seems to be a negative correlation between disorientation and transmission frequency. Next, for twist GBs, there appears to be an energy barrier around 950 mJ/m\textsuperscript{2}, above which transmission no longer occurs. No obvious correlations exist between the event stress or mean max TF and transmission frequency. As expected, transmission seems to favor smaller magnitudes of RBV, especially for twist GBs.}
\label{fig:trans_freq_vs_dis_and_eng}
\end{figure*}

\subsubsection{Frequency of Transmission}
In Figure \ref{fig:trans_freq_vs_dis_and_eng}, the frequency of transmission for each GB is plotted against the attributes considered in an effort to elucidate relationships that might correlate with transmission. Tilt and twist GBs have been shown to affect the dislocation-GB interaction differently \cite{Sangid:2012, Chandra:2017}, therefore the relationships are plotted for the two types in separate graphs. The size of each of the markers is proportional to the number of interactions used to calculate the frequency of transmission for that GB. Since there are multiple interactions for each point, we simply plot the average value of each, e.g. average event stress from each of the interactions of dislocations with that GB and average value of all the miminum RBV from each of the interactions, etc. While there do not appear to be any strong trends, there are subtle trends that would suggest disorientation angle, static GB energy, and the GB type (i.e., tilt vs twist) may be distinguishing attributes in transmission of dislocations through GBs. First, for GBs with twist or tilt about the [100] axis, the frequency of transmission appears to be negatively correlated with the disorientation angle, though there are GBs with high transmission frequency at high disorientation angles. Second, for twist GBs, the existence of a GB energy barrier to transmission seems to affect the ability of dislocations to transmit. Transmission frequency is typically high for GBs with a static energy less than 950 mJ/m\textsuperscript{2} at which point there is a steep drop in transmission frequency such that no transmission occurs for twist GBs with energy above 950mJ/m\textsuperscript{2}. No such barrier is readily visible for the tilt GBs. To determine whether or not this is due to the GB energy or the different type of GB would require further research.
 Third, there is a surprising lack of correlation between the TF and the frequency of transmission. It was expected that as the alignment of incident and transmitted slip systems increases, corresponding to a larger TF, the propensity to transmit would increase. There appears to be no such trend for twist or tilt GBs. Similarly, it was anticipated that transmission would occur more readily with a smaller available RBV, which in this case may be true. Finally, no apparent relationship between the event stress and transmission frequency is easily discernible for either twist or tilt GBs. Reasons for this lack of correlation are explored later.

\subsubsection{Predicted Slip System}
Although one cannot definitively predict the likelihood of transmission using the TF and RBV, these two attributes prove to be very effective at predicting the slip system of the transmitted dislocation. This is in accordance with earlier mentioned studies involving smaller numbers of dislocation-GB interactions \cite{Shen:1988,Sangid:2012,Abuzaid:2012,Lim:1985,Koning:2003, Li:2009,  Lee:1990, Koning:2002, Kacher:2012}. 

In the cases of transmission observed in this work, 70 full dislocations and 16 partial dislocations were observed. If the slip system predicted to emit is the one with the the maximum TF, then the correctness depends on whether full and partial dislocations are included in the list of potential outcomes. Predicting the slip system of all 86 full and partial dislocations using a potential list of 24 full and partial dislocation slip systems results in an accuracy of 55.8\% (48/86). However, if all 86 are predicted using only a potential list of 12 full dislocation slip systems the accuracy increases to 67.4\% (58/86). Finally, if only the 70 full dislocations are considered and the potential list only includes 12 full dislocations, the accuracy is 82.9\% (58/70). 

We learn some important lessons from these different levels of accuracy. Trying to predict the emitted slip system of a transmitted dislocation is not great when one attempts to include both full and partial dislocations. The partial dislocations provide so many additional options which are not likely to be selected, some of which may even have a higher TF than the full dislocations. Therefore, their inclusion initially leads to low accuracy predictions (see Table \ref{table:dislocation 36.6 RBVs and TFs}). If we predict only full slip systems but compare the prediction against the emitted full and partial dislocation slip systems, we can never get to 100\% accuracy. This is a result of attempting to predict full dislocations for some interactions where the dislocation transmitted onto a partial slip system. Since full dislocations dominate the simulations, it makes sense to predict only full slip systems for simulations that actually transmit full dislocations. In these cases, the accuracy is noteworthy. 

\begin{figure*}[t]
\centering\begin{subfigure}{0.45\linewidth}
\includegraphics[width= 1\textwidth]{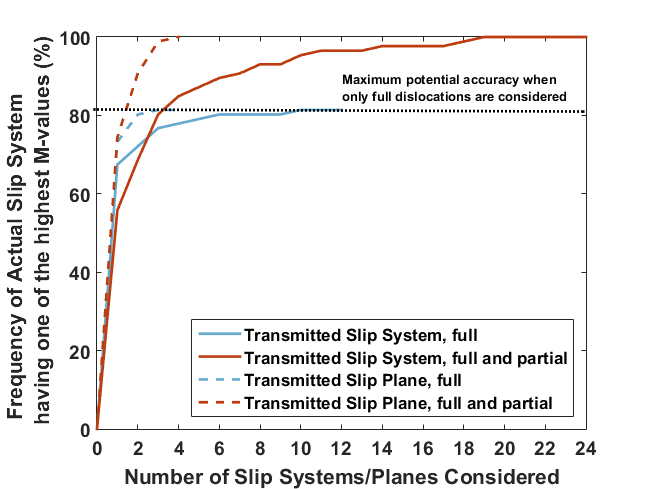}
\caption{}
\label{fig:MaxTF_accuracy}
\end{subfigure}
\begin{subfigure}{0.45\linewidth}
\centering
\includegraphics[width= 1\textwidth]{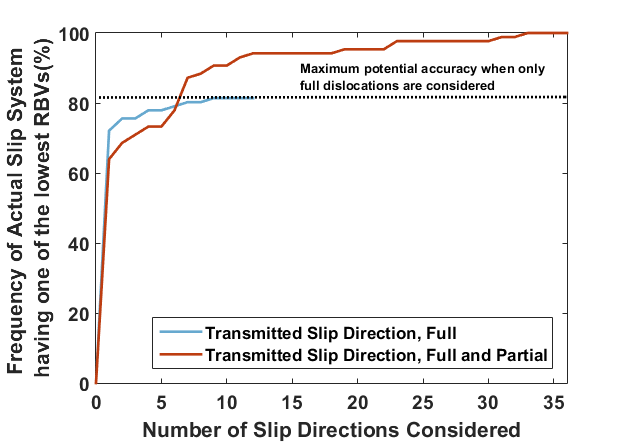}
\caption{}
\label{fig:MinRBV_accuracy}
\end{subfigure}
\caption{Ability of the a.) TF and b.) RBV to correctly predict the transmitted slip system/plane and slip direction respectively. The frequency is displayed as a function of how many slip systems/planes that produce the highest TF-values or slip directions that produce the lowest RBVs are considered for the prediction to be considered as correct. The frequency increases as additional slip systems/planes and slip directions are considered correct. The dotted black line indicates the maximum frequency capable of being achieved when only full dislocations are considered in the calculations, which is less than 100\% because 16 of the 86 transmitted slip systems were partials.}
\label{fig:MaxTF_and_MinRBV_accuracy}
\end{figure*}

In many cases, the TF of the slip system which actually transmits is the second or third highest predicted slip system according to TF. However, in many cases, the top three or so TFs are similar in magnitude, with the main difference typically being a result of a different slip direction rather than a different slip plane; this can be seen in Table \ref{table:dislocation 36.6 RBVs and TFs} where the top four dislocations according to TF have only two unique slip planes and have a range of 0.075. To illustrate how the ability to correctly select the slip system improves as more slip systems are considered acceptable, we plot the frequency of correct predictions in  Figure \ref{fig:MaxTF_accuracy}. Intuitively, the predictive capability of the TF would increase as the stipulation that the predicted transmitted slip system maximize the TF is relaxed to include the top two (or more) slip systems with the largest values of TF. This trend is shown by the solid lines in Figure \ref{fig:MaxTF_accuracy}. Here there are two solid lines, blue for predictions of full dislocations only, and red for predictions of full and partial dislocations. It should be noted that because of the 16 dislocations that emit on partial slip systems, the full slip system predictions can never get higher than 81.4\%. Interestingly, the prediction capability using the 24 full and partial slip systems increases rapidly as more slip systems are considered, reaching a frequency of 80.2\% when the prediction is considered correct if the actual TF is one of the top three values of TF possible; this is likely a sign that the full dislocations that actually transmit are simply near the partial slip systems that are predicted as in Table \ref{table:dislocation 36.6 RBVs and TFs}, so once we consider up to top three possible slip systems, it is likely to encompass the full slip system on which the dislocation transmitted.

Although the TF is reasonably accurate in predicting the correct slip system in the event of a transmitted dislocation, its accuracy is higher if we are only concerned with predicting the slip plane on which the dislocation transmits. By ignoring the slip direction and assigning the maximum TF for each unique slip plane to all slip directions on the same slip plane, the accuracy increases to 73.3\% (63/86) if only full slip systems are considered and 74.4\% (64/86) if both full and partial slip systems are considered. This is demonstrated by the dotted lines in \ref{fig:MaxTF_accuracy}. Since there are only four unique slip planes for FCC material, these lines reach their maximum once the four slip planes are considered. 

Since the accuracy of prediction goes up as additional slip systems that may be near the ``optimal'' slip system are considered, a separate analysis is done to determine how close in magnitude the TF of the actual transmitted dislocation is to the maximum TF available for each given interaction. It is found that in 62 of the 86 cases of transmission, and when considering only full dislocations, the actual transmitted dislocation has a TF with a magnitude within 20\% of the value of the maximum possible TF. However, this frequency increases to 70/86 occurrences if both full and partial dislocations are considered. This kind of information may be relevent in continuum models \cite{Lim:2011} that rely on the calculation of the obstacle stress according to equation (\ref{eqn:Tau_obs}); by knowing that the correct TF is often within 20\% of the maximum possible TF, a range for TF could be used in the calculation of the obstacle stress for a given GB.

While the slip plane can be predicted using the TF, the slip direction is similarly predicted by minimizing the RBV. A smaller RBV means that the disorder left behind in the GB is reduced for a transmission event and therefore the energy can be minimized as well. A similar analysis to the TF just discussed is repeated for the minimum predicted RBV of potential slip systems. 
 
As expected, the solid light blue line in Figure \ref{fig:MinRBV_accuracy} shows that when only full dislocations are considered, the RBV effectively predicts the transmitted slip direction in 72.1\% (62/86) of the cases of transmission. When more partial slip directions are also included, the accuracy decreases to 64.0\% (55/86), shown by the solid dark red line in Figure \ref{fig:MinRBV_accuracy}. Once again, predictions using only full dislocation slip directions can only ever achieve an accuracy 81.4\% due to the fact that some events transmitted onto partial slip systems. Similar to TF, the ability of the RBV to predict the transmitted slip direction improves as other slip directions close to the minimum are considered.

\section{Discussion}
\label{sec4}
The discussion of the results is organized into 3 sections. First, we examine criteria to predict the slip systems of transmitted dislocations. Second, we analyze attributes and their correlations with likelihood of observing a transmission event. Third, we employ machine learning to find correlations associated with different types of dislocation-GB interactions and their ability to predict the resulting event.

\subsection{Prediction of Transmission Slip Systems}
The survey of dislocation-GB interactions confirms the trends related to transmission slip systems that others have seen in regards to the geometric attributes TF and RBV. First, Shen et al.~found that the transmitted slip system could be correctly predicted in three of five experimental cases of transmission using just the TF \cite{Shen:1988}. Consistent with these experimental results, this study also shows that the TF presents an effective way to predict the slip system of a transmitted dislocation, and is even more effective in predicting just the slip plane (Figure \ref{fig:MaxTF_accuracy}). Furthermore, this study, performed on many more dislocation-GB interactions, also suggests that predictions using the maximum TF will never achieve an accuracy of 100\% because transmission on other slip systems with good alignment are consistently observed. Second, this study confirms the research done by others which shows that minimizing the RBV allows one to predict the correct slip direction for transmitted dislocations in most cases, though again, not at an accuracy of 100\% \cite{Sangid:2012,Abuzaid:2012,Lim:1985,Koning:2003,Li:2009,  Lee:1990,  Koning:2002, Kacher:2012}.

The importance of geometric criteria used to predict the slip systems involved in both experimental work and simulations is emphasized by the accuracy of the TF and the RBV. However, Table \ref{table:dislocation 36.6 RBVs and TFs} and Figure \ref{fig:MaxTF_and_MinRBV_accuracy} show that such geometric criteria are more capable when only considering full dislocations since the predictive capabilities of TF and RBV initially decrease when partial slip systems are also considered. This observation may have little effect if using RBV and TF to predict the slip system in metals with a high stacking fault energy, such as aluminum, where partial dislocations are rarely observed \cite{Nabarro2002}. In such materials, it would be easily justifiable to exclude any partial dislocations from predictions made for the transmitted slip system. By shifting the analysis to only consider the 55 full dislocations that transmitted onto full slip systems (see the far left bar in Figure \ref{fig:Classification_emitted_dislocations}), the frequency of the maximum TF predicting the correct slip plane improves to 89.1\% (49/55) and the frequency of the minimum RBV predicting the correct slip direction similarly improves to 92.7\% (51/55).

\subsection{Predicting Transmission Events}
Determining the attributes that influence a dislocation's ability to transmit through a GB is required for robust mesoscale modeling. As such, this section first describes the different trends observed that are associated with being able to predict transmission and then discusses potential complications in finding expected trends regarding the RSS.

First, as discussed in the Results, it is found that the frequency of transmission is higher for dislocations that interact with a GB that has a low static GB energy and a low disorientation angle (Figure \ref{fig:trans_freq_vs_dis_and_eng}). It is possible that GB energy is the cause while disorientation angle is correlated with GB energy. For example, for both twist and tilt GBs, there exists a GB which has high frequency of transmission despite it having the highest disorientation angle of the simulated twist or tilt GBs. However, these two high disorientation and high transmission frequency GBs also have a relatively low GB energy. Additional data would be needed to confirm this assertion.

\begin{figure*}[t]
\centering
\begin{subfigure}{0.45\linewidth}
\includegraphics[width=1\linewidth]{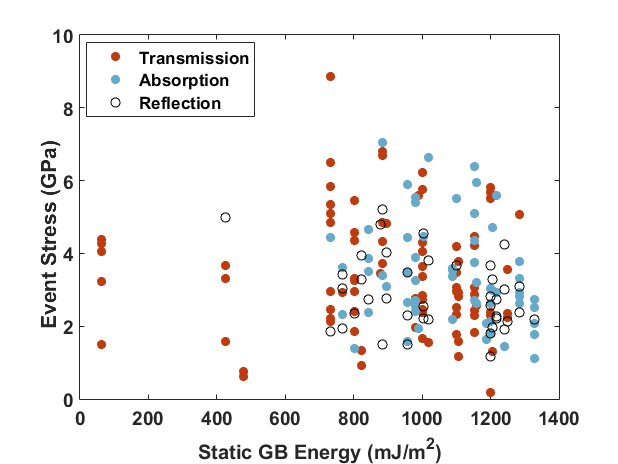}
\caption{}
\label{fig:Event_Stress_vs_GBeng}
\end{subfigure}
\begin{subfigure}{0.425\linewidth}
\centering
\includegraphics[width=1\linewidth]{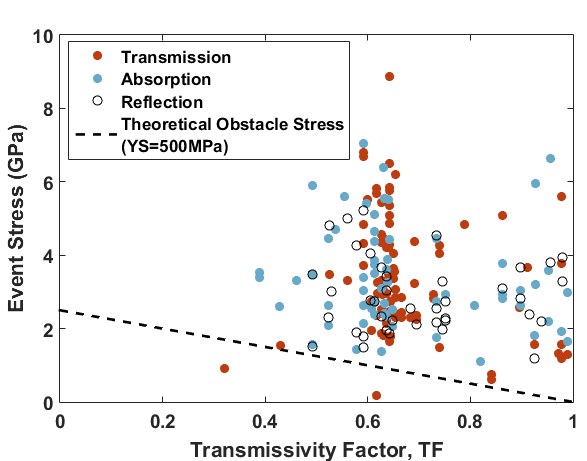}
\caption{}
\label{fig:Event_Stress_vs_TF}
\end{subfigure}
\caption{The event stress for each dislocation-GB interaction as a function of a.) the static GB energy and b.) the TF. In b.) the dotted black line indicates the theoretical obstacle stress assuming a macroscopic yield stresngth of 500 MPa. As can be seen, nearly every dislocation-GB interaction experienced an event stress that far exceeded the theoretical obstacle stress.}
\end{figure*}

This result initially seems in conflict with the work done by Sangid et al.~that found the energy barrier to transmission is negatively correlated to the GB energy \cite{Sangid:2011}. He proposes that a more ordered and stable GB with lower interfacial energy provides a stronger barrier to slip transmission and nucleation. However, this study does not attempt to measure the energy barrier to transmission, but rather the frequency of transmission correlated with the GB energy. Although a low energy GB may have a higher energy barrier, this could be interpreted as requiring a higher stress for transmission and not necessarily mean the GB is more or less likely to allow things through. It is worth noting that in Sangid's work, the highest energy barrier for transmission was about 4 times higher than the lowest barrier, but that the highest GB energy was more than 10 times higher than the lowest GB energy. Thus, the trend is not linear and a small change in GB energy would require an even smaller increase or decrease in barrier energy. To compare with this possible scenario, Figure \ref{fig:Event_Stress_vs_GBeng} plots the event stress as a function of the static GB energy for the different event types. As can be seen, no trend is immediately obvious. Thus, at this point, it is not clear why low energy GBs would simultaneously exhibit a high energy barrier to transmission (as demonstrated by Sangid) while allowing more dislocations to transmit.

Second, the influence of the GB type (twist vs tilt) and the disorientation angle on a particular GB structure may also provide an explanation for why some dislocations transmit and others do not.~Twist GBs reflected dislocations more often than tilt GBs (25\% vs 19\%) and transmitted less than tilt GBs (42\% vs 47\%).
This difference could be due to the orthogonal network of dislocations found within the twist GBs which are more dense and could offer a more significant physical barrier to transmission than the linear array of dislocations present in tilt GBs \cite{Chandra:2017, Sutton:1983}. Furthermore, the propensity to reflect increases slightly as the disorientation angle between the two grains increases. Based on the previous observations, this correlation would be expected as the density of the dislocation network increases with increasing disorientation for the GBs studied here. Li et al. also suggested that the difficulty for a dislocation to transmit is directly related to the misorientation angle \cite{Li:2009}, agreeing with the results seen in this study (c.f. Figure \ref{fig:trans_freq_vs_dis_and_eng}). Chandra et al.~also found that twist GBs offered significant resistance to transgranular crack growth, confirming the understanding that twist GBs are more resistant to such crack growth as compared to tilt GBs \cite{Chandra:2017}. 

%

Third, as demonstrated by Dewald et al.~and Bachurin et al., the actual location of the interaction between the dislocation and the GB affects the ability of the dislocation to transmit \cite{Bachurin:2010, Dewald:2007:Screw}. This suggests that the dislocation network within a GB results in regions of the GB that could present stronger physical barriers than others. This difference can lead to different outcomes for seemingly identical interactions. For example, in a simulation of the 22.62\textdegree~[100] twist GB two dislocations of the same [11$\bar{1}$](101) slip system impact the GB approximately 80\AA~apart, yet one readily transmits and the other eventually reflects off the GB. To more fully understand to what extent the location of impact and the associated atomic structure of the GB influence the interaction, a more detailed analysis would be required. Morvec et al. have shown that common geometric criteria, especially the alignment between slip planes in neighboring grains, used for the prediction of the results of dislocation-GB interactions does not always hold as the resulting event also depends on how the atoms rearrange at the event site \cite{Mrovec:2009}. While we do not dispute this fact and this may account for some of the erroneous predictions, the fact that the geometric attributes hold as well as they do, in spite of the many different atomic arrangements in all the simulations, is noteworthy.

At the outset of this work, it was hypothesized that transmission events would be correlated with the stress present on the dislocation, which would be reasonable considering Sangid's observations \cite{Sangid:2011, Sangid:2012}. However, such trends remained elusive as indicated by Figure \ref{fig:Event_Stress_vs_GBeng}. In his mesoscale model, Wagoner also proposed a similar prediction of transmission stress, given in equation \ref{eqn:Tau_obs}, but based his on the alignment of slip systems for a potential transmission event. 
To demonstrate this, the event stress for each interaction is plotted as a function of the maximum TF available for the interaction in Figure \ref{fig:Event_Stress_vs_TF}. Equation \ref{eqn:Tau_obs} is plotted as a dotted line (for a yield strength of 500 MPa), and according to Wagoner's model, stresses in excess of this value would be sufficient to cause transmission. It is clear that transmission stresses cannot be predicted so easily as nearly every point, regardless of the event that occurs at the GB, exceeds the theoretical obstacle stress for the given TF. Furthermore, there is no general trend of event stress with TF or static GB energy. 

It is possible that the nature of these simulations complicates the ability to see trends in the recorded stresses. For example, the short simulation times, and corresponding high strain rates, required for molecular dynamics simulations means that driven systems can behave differently than thermally activated systems. If not driven at high stresses, thermal fluctuations over a long time period may only be able to access a preferred event. But driven systems at high stresses may have thermal fluctuations that can access not only a preferred event, but a number of newly accessible events as well. For example, a recent study has shown that a different deformation mechanism can occur in simulations at a slower strain rate of 10\textsuperscript{6}s\textsuperscript{-1}: slip on lower Schmid factor slip systems \cite{Dupraz:2018}. Thus, if it were possible, lower strain rate simulations might observe different sets of events at different stresses, which might then exhibit a correlation in stresses not observed in the present work.

Finally, the process of transmission is likely not deterministic and more attention should be given to models that account for this fact. Modeling the process in a stochastic manner may allow a more accurate representation of activated slip systems observed here by allowing the occasional poorly aligned slip system to be activated upon transmission. Others have already applied this concept in other aspects of deformation in a quantized crystal plasticity finite element model \cite{Li:2009:Stochastic} or in the modeling of twin nucleation and/or transmission in HCP metals using a viscoplastic self-consistent model \cite{Chelladurai:2018, Niezgoda:2014, Beyerlein:2011}.

\subsection{Machine Learning Dislocation-GB Interactions}

One of the major shortcomings of the criteria used to predict the slip system of a transmitted dislocation, which is made more apparent in this study, is their current inability to predict \textit{if} a dislocation will transmit or not. For example, based on the previously discussed criteria, it is not entirely clear  as to why a dislocation would prefer to reflect over transmitting if the geometrical criteria almost always favors transmission. Furthermore, as seen in Figure \ref{fig:RSS_RBV_TF_GBeng_Diso_histograms}, each criterion, when individually considered, reveal little about the interaction event. However, machine learning processes provide promising potential for the extraction of correlations between multiple attributes that aren't readily seen. Therefore, two different analyses are performed using the WEKA machine learning software \cite{Witten:2016}. First, a decision tree is used to create a predictive model for transmission. Second, the determination of which attributes play the most significant role in determining whether or not a dislocation transmits is made by finding the gain ratio for each considered attribute.

Three separate analyses were performed on the data set to predict transmission, absorption, and reflection. It was found that the analyses to predict transmission and absorption reveal similar trends whereas the analysis to predict reflection was unsuccessful due to only a marginal increase in predictive capabilities given the attributes considered in this study. 

%
\begin{figure*}[t]
\includegraphics[width=0.8\textwidth]{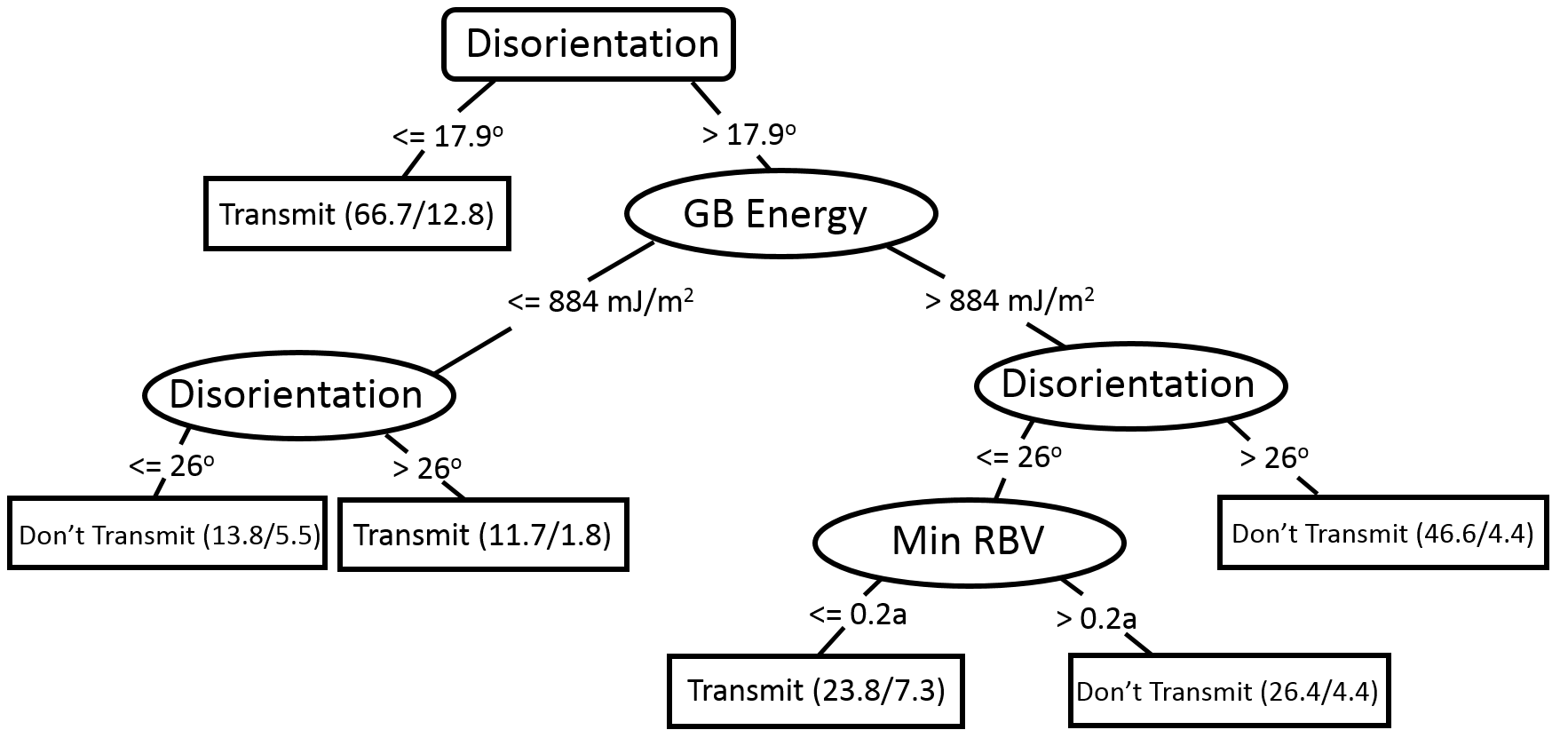}
\centering
\caption{J48 decision tree starting at the root (box with rounded corners) which produces 4 branches (ellipses) and ends with 6 leafs (boxes). The fraction inside each leaf is the number of instances that reached the leaf over how many of those instances were incorrectly classified (e.g., (46.6/4.4) means 46.6 instances made it to that branch and 4.4 were incorrectly classified where the decimals are a result of the class balancing).}
\label{fig:J48}
\end{figure*}

\subsubsection{Decision Tree Prediction of Transmission Events}
For this study, we employ the J48 method within the AttributeSelectedClassifier in WEKA which produces a simple-to-understand decision tree in order to predict a certain outcome (in this case transmission) given certain input attributes by evaluating at each branch the most important attribute of the remaining data. The J48 process is explained in more detail in \cite{Orme:2016} and \cite{Quinlan:1993}. The attributes used in the creation of the decision tree included numeric values for disorientation angle, static GB energy, maximum TF, minimum RBV, and the event stress. Attributes with nominal values for the GB type, i.e., twist vs. tilt, and for the incident dislocation type, i.e., full vs. partial, were also included. The minimum leaf size in the tree is set to 10 (meaning that each split will contain at least 10 instances), which ensures that the tree is not over constrained. In order to avoid obtaining an artificially high accuracy due to an uneven distribution of the event being predicted, class balancing is used. The GainRatioAttributeEvaluator method in WEKA is used to produce a tree based on the most relevant attributes for this study. The standard 10-fold cross validation technique is used (meaning the data is split into 10 parts, i.e., folds, then trained on 9 parts and tested on the remaining part, repeating so that each part is tested once). This produces a decision tree whose accuracy is the average of the 10 different tests. To further reduce the potential for a particular split to produce incorrectly high or low prediction accuracies, the cross-validation technique was repeated 10 times, each time using a new seed to randomly split the data differently. Using this technique, i.e., using 10-fold cross validation 10 times and averaging their accuracies, we produce the J48 decision tree shown in Figure \ref{fig:J48}, which has an average accuracy of 75.5\% with a standard deviation of 1.65\%. By comparing this result to the baseline accuracy (i.e., making the prediction based solely on the most popular outcome) of 47.7\% we find that the J48 decision tree improves the prediction of transmission vs. non-transmission by nearly 30\%. Other techniques, such as Random Forrest, could be used to improve the accuracy, but such black box methods do little to give insight into what attributes are affecting its decision process. Intuitively, the level of importance of each attribute included in the tree decreases as one travels down the decision tree. 

Several observations of the tree should be noted. First, of the seven attributes included in the creation of the tree, only three of them appear in the resulting tree. In order of importance, these attributes are disorientation angle, static GB energy, and minimum RBV. The significance of the disorientation angle is emphasized by its appearing in the tree multiple times. Second, the tree is relatively clean with only six leaves and four branches needed to predict transmission. Further measures could be made to reduce the size of the tree and simplify the results at the sacrifice of accuracy. Alternatively, accuracy could be improved by allowing a smaller leaf size, but this might over constrain the tree making the result too specific to certain results in the dataset instead of a result that is general to the majority of the data.

Biases enter the model as a result of the uneven distribution of attributes tested, as seen in Figure \ref{fig:trans_freq_vs_dis_and_eng} where, for example, certain GB energies or disorientation angles have more data points available. Despite this bias, a number of insights are gained from the J48 analysis in determining which of the investigated attributes most prominently affect transmission. First, the decision tree visually reinforces the earlier observation that, in general, GBs with a lower disorientation angle allow for transmission to occur more frequently. This is consistent with the findings by others \cite{Lim:1985,Chandra:2015,Dahlberg:2017,Davis:1966} and further supports the discrimination between high angle and low angle grain boundaries, which is commonly believed to occur at a disorientation angle of 15\textdegree  \cite{Gottstein:2013}. In fact, as seen by the first branch of the decision tree, for GBs with a disorientation angle of less than 18\textdegree, none of the other attributes have an affect on transmission. It should be noted that there are only 9 GBs with a disorientation angle less than 18\textdegree. However, for some GBs that have higher disorientation angles, above 18\textdegree, a lower static GB energy is preferable for transmission. Second, as expected, transmission is predicted to occur more frequently when the minimum available RBV is smaller. Finally, for dislocations that interact with a GB that has high energy and a disorientation angle above 26\textdegree, transmission is never predicted.

\begin{table}
\centering
 \caption{Gain Ratio for each of the considered attributes for predicting transmission. Not surprisingly, minimum RBV and disorientation angle provide the most information.}
 \label{table:Information_gain_table}
\small
 \begin{tabular}{@{}*{2}{c}@{}} 
 \hline
 \thead{Attribute} & \thead{Transmission\\Gain Ratio}\\ [0.5ex] 
 \hline
Disorientation Angle & 0.1675\\
Minimum RBV & 0.1482\\
Static GB Energy & 0.1314\\
Max Transmissivity Factor & 0.0756\\
Partial vs. Full & 0.0070\\
Twist vs. Tilt GB & 0.0018\\
Event Stress & 0.0000\\
 \hline
 \end{tabular}
\end{table}

\subsubsection{Attribute Evaluation of Transmission Events}
WEKA is also capable of determining the relative importance in predicting the defined class for each included attribute by using the Gain Ratio Attribute Evaluator. This function measures the amount by which each attribute decreases the overall entropy \cite{Sharma:2012}. Attributes which result in a larger decrease in entropy, or reduce the uncertainty in the outcome, are said to provide more information. An important distinction that should be made between this method and the J48 decision tree is that the Gain Ratio Attribute Evaluator considers the whole data set when calculating the gain ratio for each attribute. This is different from J48 in that the J48 method only considers the data available at each node in the branch to determine which is the most important attribute at that point. The gain ratio for each attribute is displayed in Table \ref{table:Information_gain_table}. This result agrees well with the decision tree created earlier in that the top three attributes according to their gain ratio are the three attributes present in the tree. Furthermore, the importance of the disorientation angle is reinforced by its being both the root of the tree and the most informative attribute. Finally, it is shown that in this study, the event stress does not contribute at all to predicting whether or not transmission will occur as it has a gain ratio of 0.0.

\subsubsection{Machine Learned Attributes Affecting Transmission Events}
The machine learning process provides unique insight into the transmission process. It is significant that such an accurate model to predict transmission can be created using predominately geometric attributes, especially RBV and disorientation. The machine learning is not able to find any correlations to event stress. The question then is whether this is a result of the approach used in this work as discussed above, or if the stress simply plays a secondary role to other attributes like the prominent geometric criteria. However, that is not to say that improvements to the model could not be made. By including a larger variety of attributes in the transmission model created, the ability to predict transmission improves. Perhaps there are other attributes not included in this study, such as temperature, that, if included in the model, could further improve its ability to predict transmission. Therefore, the value of this model lies in demonstrating the potential of more thorough studies which consider more attributes to produce a superior model to predict transmission.

\subsubsection{Machine Learning of Absorption/Reflection}
The same procedures used to create the decision tree and gain ratio table for predicting transmission are performed for the creation of decision trees to predict absorption and reflection of dislocations as well as to determine which attributes were most informative of the subsequent GB event. For brevity, the results of this analysis are discussed here and the resulting J48 decision trees and gain ratio tables are included in the supplemental material (Figures S4 and S5 and Tables S3 and S4). 

Similar to transmission, absorption of a dislocation is correctly predicted 77.0\% of the time with a standard deviation of 1.29\% using a J48 decision tree, an improvement of about 30\% from the baseline accuracy of 47.6\%. Unsurprisingly, the absorption decision tree reinforces the trend found for transmission that disorientation angle plays a significant role in the event. Here, a higher disorientation angle is found to be the best indicator that a dislocation will absorb rather than reflect or transmit. Furthermore, the gain ratio table for absorption is similar in order of the more influential attributes to that of the attributes for transmission, with disorientation angle and minimum RBV being the two most informative attributes. The trends displayed here are nearly equivalent to those found in the transmission tree and information table, confirming that transmission prefers smaller minimum RBVs and lower disorientation angles.

In contrast to transmission and absorption, reflection of a dislocation was unsuccessfully predicted using machine learning. The J48 decision tree for reflection improved the baseline accuracy by only 11.5\% , from 49.7\% to 60.2\% with a standard deviation of 3.2\%. After considering the decision tree and the gain ratio for each of the given variables, this is not surprising. Unlike the other two decision trees, the decision tree to predict reflection has nine branches and 11 leaves, meaning it has many more frequent and smaller splits of the data, resulting in an over-constrained tree. Furthermore, the root of the tree and the highest gain ratio value of any variable used to predict reflection is the event stress, which is already known to contain some uncertainty. Finally, none of the other variables associated with transmission or absorption are correlated with reflection, casting doubt on the ability to create an effective predictive model of reflection with the current data set. Significantly more data is required to understand what attributes influence reflection of dislocations at grain boundaries.

\section{Conclusion}
\label{sec5}

This study utilizes molecular dynamics to contribute new insights into dislocation-GB interactions through the study of numerous interactions occurring in a large variety of Ni bicrystals. Geometric attributes as well as stresses and energies are used to characterize the interactions. The major goal of the study is to understand and ultimately predict whether a given dislocation will transmit through a GB and if so, onto what slip system.  We find, as others do, that dislocation-GB interactions are an extremely complex process. Despite this inherent difficulty, the following general conclusions can be made.

\noindent \emph{Prediction of Transmitted Slip Systems}

    \begin{enumerate}
        \item  Transmissivity Factor (TF) predicts with reasonable accuracy the slip system for transmission; an accuracy of 67.4\% is obtained if only considering transmitted full dislocations and 55.8\% if transmitted partial dislocations are also considered. 
        \item The accuracy of the TF improves if used to only predict the slip plane of the transmitted dislocation, 73.3\% correct if only considering transmitted full dislocations and 74.4\% accurate if transmitted partials are also considered. 
        \item For the majority of transmitted dislocations, the TF is within 20\% of the maximum TF capable for the given dislocation-GB interaction.
        \item RBV predicts well the slip direction for transmission. It correctly predicts the slip direction 72.1\% of the time when only full transmitted dislocations are considered, decreasing to 64.0\% when transmitted partials are included. 
        \item Predicting the slip system of a transmitted dislocation is more accurate when only considering full transmitted dislocations as possible transmitted dislocations. Past studies have focused solely on full dislocations \cite{Abuzaid:2012, Shen:1988, Livingston:1957, Dahlberg:2017}, though their reasoning for doing so is not discussed, but it may be due to the fact that they observed very little partial slip activity.
        \item The TF can be used to predict the correct slip plane 89.1\% of the time and the RBV can be used to predict the correct slip direction 92.7\% of the time for a transmitted dislocation when only accounting for full incident dislocations that transmitted as full dislocations.
        \item While no correlation betweeen event stresses of dislocation-GB interactions appear for the entire dataset, some consistency in event stresses for the same slip system over several GBs is observed. 
    \end{enumerate}
    
    \noindent \emph{Prediction of Dislocation-GB Events}
        
    \begin{enumerate}
        \item The expected geometric trends in regards to transmission, e.g. increased transmission frequency for smaller disorientation angles, hold for a large variety of GBs.
        \item Reflection of dislocations occurs more frequently at twist GBs than tilt GBs while the opposite is true of transmission. Both twist and tilt GBs are equally likely to absorb dislocations.
        \item Partial dislocations are more likely to transmit than full dislocations; 53.5\% of partial dislocations transmit and 43.2\% of full dislocations transmit.
        \item Twist GBs appear to have a GB energy barrier to transmission; approximately 950 mJ/m\textsuperscript{2} for the GBs studied here. 
        \item Utilizing machine learning to create a simple J48 decision tree, transmission can be correctly predicted 75.5\% of the time and absorption can be correctly predicted 77\% of the time. Reflection is not effectively predicted in this study.
        \item The relative importance of the studied attributes influence on the interaction event is provided, also confirming that transmission favors low-angle GBs. 
        \item Improvements to predictive capabilities can be achieved by including more attributes in the model. This demonstrates the potential for a physics based model that can predict transmission/absorption.
        \item Transmission can be reasonably predicted by knowing just the dislocation type  of the incident dislocation (full vs partial) and the minimum RBV possible. 
        \item Although this study reveals important geometrical relationships between transmission and other dislocation-GB events, it does not do a good job of measuring stress and does not reveal conclusive relationships between RSS and the resulting event. 
        
            However, Figure \ref{fig:Event_Stress_vs_dis_vs_SS} reveals the potential for elucidating relationships between the incident slip system, the disorientation angle, and the given stress. A more carefully controlled study of the stresses involved, like \cite{Wyman:2017}, would need to be conducted for such relationships to be revealed.
 
    \end{enumerate}

Although there remains much work to be done to fully understand dislocation-GB interactions, the current work offers new insights into attributes that affect the transmission of dislocations and the potential challenges in more accurately modeling such interactions.

\section*{Acknowledgements}
\label{sec6}
This work was funded by the U.S. Department of Energy grant number DE-SC0012587. The author also wishes to acknowledge the support provided by Ricky Wyman and Adam Herron and their vital contributions in the early stages of the project.

\section*{Data Availability}
\label{sec7}
The raw data required to reproduce these findings is not archived due to its inherently large size. The processed data required to reproduce these findings are available to download from \url{http://dx.doi.org/10.17632/m7k9fzwyyr.1}.

 \bibliographystyle{elsarticle-num}

\bibliography{Research_Paper_Bib}

\begin{thebibliography}{10}
\expandafter\ifx\csname url\endcsname\relax
  \def\url#1{\texttt{#1}}\fi
\expandafter\ifx\csname urlprefix\endcsname\relax\def\urlprefix{URL }\fi
\expandafter\ifx\csname href\endcsname\relax
  \def\href#1#2{#2} \def\path#1{#1}\fi

\bibitem{Farkas:2013}
D.~Farkas, Atomistic simulations of metallic microstructures, Current Opinion
  in Solid State and Materials Science 17~(6) (2013) 284--297.
\newblock \href {http://dx.doi.org/10.1016/j.cossms.2013.11.002}
  {\path{doi:10.1016/j.cossms.2013.11.002}}.

\bibitem{Kacher:2014}
J.~Kacher, B.~P. Eftink, B.~Cui, I.~M. Robertson, Dislocation interactions with
  grain boundaries, Current Opinion in Solid State and Materials Science 18~(4)
  (2014) 227--243.
\newblock \href
  {http://dx.doi.org/http://dx.doi.org/10.1016/j.cossms.2014.05.004}
  {\path{doi:http://dx.doi.org/10.1016/j.cossms.2014.05.004}}.

\bibitem{Hasnaoui:2004}
A.~Hasnaoui, P.~Derlet, H.~Van~Swygenhoven, Interaction between dislocations
  and grain boundaries under an indenter--a molecular dynamics simulation, Acta
  Materialia 52~(8) (2004) 2251--2258.
\newblock \href {http://dx.doi.org/10.1016/j.actamat.2004.01.018}
  {\path{doi:10.1016/j.actamat.2004.01.018}}.

\bibitem{Voyiadjis:2016}
G.~Z. Voyiadjis, M.~Yaghoobi, Role of grain boundary on the sources of size
  effects, Computational Materials Science 117 (2016) 315--329.
\newblock \href {http://dx.doi.org/10.1016/j.commatsci.2016.01.025}
  {\path{doi:10.1016/j.commatsci.2016.01.025}}.

\bibitem{Hall:1951}
E.~Hall, The deformation and ageing of mild steel: Iii discussion of results,
  Proceedings of the Physical Society. Section B 64~(9) (1951) 747.

\bibitem{Petch:1953}
N.~J. Petch, The cleavage strengh of polycrystals, J. of the Iron and Steel
  Inst. 174 (1953) 25--28.

\bibitem{Ovidko:2018}
I.~Ovid'ko, R.~Valiev, Y.~Zhu, Review on superior strength and enhanced
  ductility of metallic nanomaterials, Progress in Materials Science 94 (2018)
  462 -- 540.
\newblock \href {http://dx.doi.org/10.1016/j.pmatsci.2018.02.002}
  {\path{doi:10.1016/j.pmatsci.2018.02.002}}.

\bibitem{Koch:2003}
C.~C. Koch, Ductility in nanostructured and ultra fine-grained materials:
  Recent evidence for optimism, in: Journal of Metastable and Nanocrystalline
  Materials, Vol.~18, Trans Tech Publ, 2003, pp. 9--20.
\newblock \href {http://dx.doi.org/10.4028/www.scientific.net/JMNM.18.9}
  {\path{doi:10.4028/www.scientific.net/JMNM.18.9}}.

\bibitem{Meyers:2006}
M.~A. Meyers, A.~Mishra, D.~J. Benson, Mechanical properties of nanocrystalline
  materials, Progress in Materials Science 51~(4) (2006) 427--556.
\newblock \href {http://dx.doi.org/10.1016/j.pmatsci.2005.08.003}
  {\path{doi:10.1016/j.pmatsci.2005.08.003}}.

\bibitem{Hansen:2017}
L.~T. Hansen, B.~E. Jackson, D.~T. Fullwood, S.~I. Wright, M.~De~Graef, E.~R.
  Homer, R.~H. Wagoner, Influence of noise-generating factors on
  cross-correlation electron backscatter diffraction (EBSD)
  measurement of geometrically necessary dislocations (GNDs),
  Microscopy and Microanalysis 23~(3) (2017) 460--471.
\newblock \href {http://dx.doi.org/10.1017/S1431927617000204}
  {\path{doi:10.1017/S1431927617000204}}.

\bibitem{Jackson:2016}
B.~E. Jackson, J.~J. Christensen, S.~Singh, M.~De~Graef, D.~T. Fullwood, E.~R.
  Homer, R.~H. Wagoner, Performance of dynamically simulated reference patterns
  for cross-correlation electron backscatter diffraction, Microscopy and
  Microanalysis 22~(4) (2016) 789--802.
\newblock \href {http://dx.doi.org/10.1017/S143192761601148X}
  {\path{doi:10.1017/S143192761601148X}}.

\bibitem{Ruggles:2013}
T.~Ruggles, D.~Fullwood, Estimations of bulk geometrically necessary
  dislocation density using high resolution EBSD,
  Ultramicroscopy 133 (2013) 8--15.
\newblock \href {http://dx.doi.org/10.1016/j.ultramic.2013.04.011}
  {\path{doi:10.1016/j.ultramic.2013.04.011}}.

\bibitem{Ruggles:2016}
T.~Ruggles, T.~Rampton, A.~Khosravani, D.~Fullwood, The effect of length scale
  on the determination of geometrically necessary dislocations via
  EBSD continuum dislocation microscopy, Ultramicroscopy 164
  (2016) 1--10.
\newblock \href {http://dx.doi.org/10.1016/j.ultramic.2016.03.003}
  {\path{doi:10.1016/j.ultramic.2016.03.003}}.

\bibitem{Bong:2017}
H.~J. Bong, H.~Lim, M.-G. Lee, D.~T. Fullwood, E.~R. Homer, R.~H. Wagoner, An
  rve procedure for micromechanical prediction of mechanical behavior of
  dual-phase steel, Materials Science and Engineering: A 695 (2017) 101--111.
\newblock \href {http://dx.doi.org/10.1016/j.msea.2017.04.032}
  {\path{doi:10.1016/j.msea.2017.04.032}}.

\bibitem{Lim:2011}
H.~Lim, M.~G. Lee, J.~H. Kim, B.~L. Adams, R.~H. Wagoner, Simulation of
  polycrystal deformation with grain and grain boundary effects, International
  Journal of Plasticity 27~(9) (2011) 1328--1354.
\newblock \href {http://dx.doi.org/10.1016/j.ijplas.2011.03.001}
  {\path{doi:10.1016/j.ijplas.2011.03.001}}.

\bibitem{Wyman:2017}
R.~D. Wyman, D.~T. Fullwood, R.~H. Wagoner, E.~R. Homer, Variability of
  non-schmid effects in grain boundary dislocation nucleation criteria, Acta
  Materialia 124 (2017) 588--597.
\newblock \href {http://dx.doi.org/10.1016/j.actamat.2016.11.005}
  {\path{doi:10.1016/j.actamat.2016.11.005}}.

\bibitem{Olmsted:2009:energy}
D.~L. Olmsted, S.~M. Foiles, E.~A. Holm, Survey of computed grain boundary
  properties in face-centered cubic metals: I. grain boundary energy, Acta
  Materialia 57~(13) (2009) 3694--3703.
\newblock \href {http://dx.doi.org/10.1016/j.actamat.2009.04.007}
  {\path{doi:10.1016/j.actamat.2009.04.007}}.

\bibitem{Sutton:1995}
A.~P. Sutton, Interfaces in crystalline materials, Clarendon Press, 1995.

\bibitem{Shen:1988}
Z.~Shen, R.~Wagoner, W.~Clark, Dislocation and grain boundary interactions in
  metals, Acta metallurgica 36~(12) (1988) 3231--3242.
\newblock \href {http://dx.doi.org/10.1016/0001-6160(88)90058-2}
  {\path{doi:10.1016/0001-6160(88)90058-2}}.

\bibitem{Sangid:2012}
M.~D. Sangid, T.~Ezaz, H.~Sehitoglu, Energetics of residual dislocations
  associated with slip-–twin and slip–-gbs interactions, Materials Science
  and Engineering: A 542 (2012) 21--30.
\newblock \href
  {http://dx.doi.org/http://dx.doi.org/10.1016/j.msea.2012.02.023}
  {\path{doi:http://dx.doi.org/10.1016/j.msea.2012.02.023}}.

\bibitem{Wang:2015}
J.~Wang, Atomistic simulations of dislocation pileup: Grain boundaries
  interaction, JOM 67~(7) (2015) 1515--1525.
\newblock \href {http://dx.doi.org/10.1007/s11837-015-1454-0}
  {\path{doi:10.1007/s11837-015-1454-0}}.

\bibitem{Clark:1989}
W.~Clark, C.~Wise, Z.~Shen, R.~Wagoner, The use of the transmission electron
  microscope in analyzing slip propagation across interfaces, Ultramicroscopy
  30~(1-2) (1989) 76--89.
\newblock \href {http://dx.doi.org/10.1016/0304-3991(89)90175-7}
  {\path{doi:10.1016/0304-3991(89)90175-7}}.

\bibitem{Abuzaid:2012}
W.~Abuzaid, M.~D. Sangid, H.~Sehitoglu, J.~Carroll, J.~Lambros, The role of
  slip transmission on plastic strain accumulation across grain boundaries,
  Procedia IUTAM 4 (2012) 169--178.
\newblock \href
  {http://dx.doi.org/http://dx.doi.org/10.1016/j.piutam.2012.05.019}
  {\path{doi:http://dx.doi.org/10.1016/j.piutam.2012.05.019}}.

\bibitem{Lim:1985}
L.~C. Lim, R.~Raj, Continuity of slip screw and mixed crystal dislocations
  across bicrystals of nickel at 573 k, Acta Metallurgica 33~(8) (1985)
  1577--1583.
\newblock \href
  {http://dx.doi.org/http://dx.doi.org/10.1016/0001-6160(85)90057-4}
  {\path{doi:http://dx.doi.org/10.1016/0001-6160(85)90057-4}}.

\bibitem{Koning:2003}
M.~de~Koning, R.~J. Kurtz, V.~V. Bulatov, C.~H. Henager, R.~G. Hoagland,
  W.~Cai, M.~Nomura, Modeling of dislocation–grain boundary interactions in
  fcc metals, Journal of Nuclear Materials 323~(2-3) (2003) 281--289.
\newblock \href {http://dx.doi.org/10.1016/j.jnucmat.2003.08.008}
  {\path{doi:10.1016/j.jnucmat.2003.08.008}}.

\bibitem{Livingston:1957}
J.~Livingston, B.~Chalmers, Multiple slip in bicrystal deformation, Acta
  Metallurgica 5~(6) (1957) 322--327.
\newblock \href {http://dx.doi.org/10.1016/0001-6160(57)90044-5}
  {\path{doi:10.1016/0001-6160(57)90044-5}}.

\bibitem{Lee:1989}
T.~Lee, I.~Robertson, H.~Birnbaum, Prediction of slip transfer mechanisms
  across grain boundaries, Scripta metallurgica 23~(8) (1989) 1467.
\newblock \href {http://dx.doi.org/10.1016/0036-9748(89)90534-6}
  {\path{doi:10.1016/0036-9748(89)90534-6}}.

\bibitem{Weygand:2002}
D.~Weygand, L.~H. Friedman, E.~V. der Giessen, A.~Needleman, Aspects of
  boundary-value problem solutions with three-dimensional dislocation dynamics,
  Modelling and Simulation in Materials Science and Engineering 10~(4) (2002)
  437.

\bibitem{Zhang:2017}
L.~Zhang, Y.~Xiang, Motion of grain boundaries incorporating dislocation
  structure, Journal of the Mechanics and Physics of Solids 117 (2018)
  157--178.
\newblock \href {http://dx.doi.org/10.1016/j.jmps.2018.05.001}
  {\path{doi:10.1016/j.jmps.2018.05.001}}.

\bibitem{Li:2009}
Z.~Li, C.~Hou, M.~Huang, C.~Ouyang, Strengthening mechanism in
  micro-polycrystals with penetrable grain boundaries by discrete dislocation
  dynamics simulation and Hall–-Petch effect,
  Computational Materials Science 46~(4) (2009) 1124--1134.
\newblock \href {http://dx.doi.org/10.1016/j.commatsci.2009.05.021}
  {\path{doi:10.1016/j.commatsci.2009.05.021}}.

\bibitem{Shen:1986}
Z.~Shen, R.~Wagoner, W.~Clark, Dislocation pile-up and grain boundary
  interactions in 304 stainless steel, Scripta metallurgica 20~(6) (1986)
  921--926.
\newblock \href {http://dx.doi.org/10.1016/0036-9748(86)90467-9}
  {\path{doi:10.1016/0036-9748(86)90467-9}}.

\bibitem{Sangid:2011}
M.~D. Sangid, T.~Ezaz, H.~Sehitoglu, I.~M. Robertson, Energy of slip
  transmission and nucleation at grain boundaries, Acta Materialia 59~(1)
  (2011) 283--296.
\newblock \href {http://dx.doi.org/10.1016/j.actamat.2010.09.032}
  {\path{doi:10.1016/j.actamat.2010.09.032}}.

\bibitem{Dewald:2007:Screw}
M.~P. Dewald, W.~A. Curtin, Multiscale modelling of dislocation/grain boundary
  interactions. II. screw dislocations impinging on tilt
  boundaries in al, Philosophical Magazine 87~(30) (2007) 4615--4641.
\newblock \href {http://dx.doi.org/10.1080/14786430701297590}
  {\path{doi:10.1080/14786430701297590}}.

\bibitem{Chandra:2015}
S.~Chandra, M.~K. Samal, V.~M. Chavan, R.~J. Patel, Atomistic simulations of
  interaction of edge dislocation with twist grain boundaries in al-effect of
  temperature and boundary misorientation, Materials Science and Engineering: A
  646 (2015) 25--32.
\newblock \href {http://dx.doi.org/10.1016/j.msea.2015.08.049}
  {\path{doi:10.1016/j.msea.2015.08.049}}.

\bibitem{Koning:2002}
M.~d. Koning, R.~Miller, V.~V. Bulatov, F.~F. Abraham, Modelling grain-boundary
  resistance in intergranular dislocation slip transmission, Philosophical
  Magazine A 82~(13) (2002) 2511--2527.
\newblock \href {http://dx.doi.org/10.1080/01418610208240050}
  {\path{doi:10.1080/01418610208240050}}.

\bibitem{Swygenhoven:2006}
H.~Van~Swygenhoven, P.~Derlet, A.~Fr{\o}seth, Nucleation and propagation of
  dislocations in nanocrystalline fcc metals, Acta Materialia 54~(7) (2006)
  1975--1983.
\newblock \href {http://dx.doi.org/10.1016/j.actamat.2005.12.026}
  {\path{doi:10.1016/j.actamat.2005.12.026}}.

\bibitem{Bachurin:2010}
D.~V. Bachurin, D.~Weygand, P.~Gumbsch, Dislocation–grain boundary
  interaction in 〈111〉 textured thin metal films, Acta Materialia 58~(16)
  (2010) 5232--5241.
\newblock \href {http://dx.doi.org/10.1016/j.actamat.2010.05.037}
  {\path{doi:10.1016/j.actamat.2010.05.037}}.

\bibitem{Mrovec:2009}
M.~Mrovec, C.~Elsässer, P.~Gumbsch, Interactions between lattice dislocations
  and twin boundaries in tungsten: A comparative atomistic simulation study,
  Philosophical Magazine 89~(34-36) (2009) 3179--3194.
\newblock \href {http://dx.doi.org/10.1080/14786430903246346}
  {\path{doi:10.1080/14786430903246346}}.

\bibitem{Lee:1990}
T.~Lee, I.~Robertson, H.~Birnbaum, Tem in situ deformation study of the
  interaction of lattice dislocations with grain boundaries in metals,
  Philosophical Magazine A 62~(1) (1990) 131--153.
\newblock \href {http://dx.doi.org/10.1080/01418619008244340}
  {\path{doi:10.1080/01418619008244340}}.

\bibitem{Plimpton:1995}
S.~Plimpton, Fast parallel algorithms for short-range molecular dynamics,
  Journal of computational physics 117~(1) (1995) 1--19.
\newblock \href {http://dx.doi.org/10.1006/jcph.1995.1039}
  {\path{doi:10.1006/jcph.1995.1039}}.

\bibitem{Hoyt:2005}
J.~J. Hoyt, S.~M. Foiles, Computation of grain boundary stiffness and mobility
  from boundary fluctuations.

\bibitem{Siegel:2005}
D.~J. Siegel, Generalized stacking fault energies, ductilities, and
  twinnabilities of ni and selected ni alloys, Applied Physics Letters 87~(12)
  (2005) 121901.
\newblock \href {http://dx.doi.org/10.1063/1.2051793}
  {\path{doi:10.1063/1.2051793}}.

\bibitem{Rice:1992}
J.~R. Rice, Dislocation nucleation from a crack tip: an analysis based on the
  peierls concept, Journal of the Mechanics and Physics of Solids 40~(2) (1992)
  239--271.
\newblock \href {http://dx.doi.org/10.1016/S0022-5096(05)80012-2}
  {\path{doi:10.1016/S0022-5096(05)80012-2}}.

\bibitem{Dahlberg:2017}
C.~F.~O. Dahlberg, Y.~Saito, M.~S. Öztop, J.~W. Kysar, Geometrically necessary
  dislocation density measurements at a grain boundary due to wedge indentation
  into an aluminum bicrystal, Journal of the Mechanics and Physics of Solids
  105 (2017) 131--149.
\newblock \href {http://dx.doi.org/10.1016/j.jmps.2017.05.005}
  {\path{doi:10.1016/j.jmps.2017.05.005}}.

\bibitem{Alexander:2010}
S.~Alexander, Visualization and analysis of atomistic simulation data with
  ovito–the open visualization tool, Modelling and Simulation in Materials
  Science and Engineering 18~(1) (2010) 015012.
\newblock \href {http://dx.doi.org/10.1088/0965-0393/18/1/015012}
  {\path{doi:10.1088/0965-0393/18/1/015012}}.

\bibitem{Xu:2016}
S.~Xu, L.~Xiong, Y.~Chen, D.~L. McDowell, Sequential slip transfer of
  mixed-character dislocations across $\Sigma$3 coherent twin
  boundary in fcc metals: a concurrent atomistic-continuum study, npj
  Computational Materials 2 (2016) 15016.
\newblock \href {http://dx.doi.org/10.1038/npjcompumats.2015.16}
  {\path{doi:10.1038/npjcompumats.2015.16}}.

\bibitem{Hamid:2017}
M.~Hamid, H.~Lyu, B.~J. Schuessler, P.~C. Wo, H.~M. Zbib, Modeling and
  characterization of grain boundaries and slip transmission in dislocation
  density-based crystal plasticity, Crystals 7~(6) (2017) 152.
\newblock \href {http://dx.doi.org/10.3390/cryst7060152}
  {\path{doi:10.3390/cryst7060152}}.

\bibitem{Aust:1954}
K.~Aust, N.~Chen, Effect of orientation difference on the plastic deformation
  of aluminum bicrystals, Acta Metallurgica 2~(4) (1954) 632 -- 638.
\newblock \href {http://dx.doi.org/10.1016/0001-6160(54)90199-6}
  {\path{doi:10.1016/0001-6160(54)90199-6}}.

\bibitem{Davis:1966}
K.~Davis, E.~Teghtsoonian, A.~Lu, Slip band continuity across grain boundaries
  in aluminum, Acta Metallurgica 14~(12) (1966) 1677--1684.
\newblock \href {http://dx.doi.org/10.1016/0001-6160(66)90020-4}
  {\path{doi:10.1016/0001-6160(66)90020-4}}.

\bibitem{Gao:2017}
Y.~Gao, Z.~Jin, Interactions between lattice dislocation and lomer-type
  low-angle grain boundary in nickel, Computational Materials Science 138
  (2017) 225--235.
\newblock \href {http://dx.doi.org/10.1016/j.commatsci.2017.06.025}
  {\path{doi:10.1016/j.commatsci.2017.06.025}}.

\bibitem{Chandra:2017}
S.~Chandra, N.~N. Kumar, M.~Samal, V.~Chavan, S.~Raghunathan, An atomistic
  insight into the fracture behavior of bicrystal aluminum containing twist
  grain boundaries, Computational Materials Science 130 (2017) 268--281.
\newblock \href {http://dx.doi.org/10.1016/j.commatsci.2017.01.023}
  {\path{doi:10.1016/j.commatsci.2017.01.023}}.

\bibitem{Kacher:2012}
J.~Kacher, I.~Robertson, Quasi-four-dimensional analysis of dislocation
  interactions with grain boundaries in 304 stainless steel, Acta Materialia
  60~(19) (2012) 6657--6672.
\newblock \href {http://dx.doi.org/10.1016/j.actamat.2012.08.036}
  {\path{doi:10.1016/j.actamat.2012.08.036}}.

\bibitem{Nabarro2002}
F.~R. Nabarro, M.~S. Duesbery, Dislocations in solids, Vol.~11, Elsevier, 2002.

\bibitem{Sutton:1983}
A.~P. Sutton, V.~Vitek, On the structure of tilt grain boundaries in cubic
  metals i. symmetrical tilt boundaries, Phil. Trans. R. Soc. Lond. A
  309~(1506) (1983) 1--36.

\bibitem{Dupraz:2018}
M.~Dupraz, Z.~Sun, C.~Brandl, H.~Van~Swygenhoven, Dislocation interactions at
  reduced strain rates in atomistic simulations of nanocrystalline al, Acta
  Materialia 144 (2018) 68--79.
\newblock \href {http://dx.doi.org/10.1016/j.actamat.2017.10.043}
  {\path{doi:10.1016/j.actamat.2017.10.043}}.

\bibitem{Li:2009:Stochastic}
L.~Li, P.~M. Anderson, M.-G. Lee, E.~Bitzek, P.~Derlet, H.~Van~Swygenhoven, The
  stress--strain response of nanocrystalline metals: a quantized crystal
  plasticity approach, Acta Materialia 57~(3) (2009) 812--822.
\newblock \href {http://dx.doi.org/10.1016/j.actamat.2008.10.035}
  {\path{doi:10.1016/j.actamat.2008.10.035}}.

\bibitem{Chelladurai:2018}
I.~Chelladurai, D.~Adams, D.~T. Fullwood, M.~P. Miles, S.~Niezgoda, I.~J.
  Beyerlein, M.~Knezevic, Modeling of trans-grain twin transmission in
  AZ31 via a neighborhood-based viscoplastic self-consistent
  model, International Journal of Plasticity\href
  {http://dx.doi.org/10.1016/j.ijplas.2018.03.012}
  {\path{doi:10.1016/j.ijplas.2018.03.012}}.

\bibitem{Niezgoda:2014}
S.~R. Niezgoda, A.~K. Kanjarla, I.~J. Beyerlein, C.~N. Tom{\'e}, Stochastic
  modeling of twin nucleation in polycrystals: an application in hexagonal
  close-packed metals, International journal of plasticity 56 (2014) 119--138.
\newblock \href {http://dx.doi.org/10.1016/j.ijplas.2013.11.005}
  {\path{doi:10.1016/j.ijplas.2013.11.005}}.

\bibitem{Beyerlein:2011}
I.~J. Beyerlein, R.~McCabe, C.~Tome, Stochastic processes of $\{$1012$\}$
  deformation twinning in hexagonal close-packed polycrystalline zirconium and
  magnesium, International Journal for Multiscale Computational Engineering
  9~(4) (2011) 459--480.
\newblock \href {http://dx.doi.org/10.1615/IntJMultCompEng.v9.i4.80}
  {\path{doi:10.1615/IntJMultCompEng.v9.i4.80}}.

\bibitem{Witten:2016}
I.~H. Witten, E.~Frank, M.~A. Hall, C.~J. Pal, Data Mining: Practical machine
  learning tools and techniques, Morgan Kaufmann Publishers, Inc., 2016.

\bibitem{Orme:2016}
A.~D. Orme, I.~Chelladurai, T.~M. Rampton, D.~T. Fullwood, A.~Khosravani, M.~P.
  Miles, R.~K. Mishra, Insights into twinning in mg az31: a combined ebsd and
  machine learning study, Computational Materials Science 124 (2016) 353--363.
\newblock \href {http://dx.doi.org/10.1016/j.commatsci.2016.08.011}
  {\path{doi:10.1016/j.commatsci.2016.08.011}}.

\bibitem{Quinlan:1993}
J.~R. Quinlan, C4. 5: Programs for machine learning, Vol.~38, Morgan Kauffmann
  Publishers, Inc., 1993.
\newblock \href {http://dx.doi.org/10.1016/C2009-0-27846-9}
  {\path{doi:10.1016/C2009-0-27846-9}}.

\bibitem{Gottstein:2013}
G.~Gottstein, Physical Foundations of Materials Science, Springer-Verlag Berlin
  Heidelberg, 2004.
\newblock \href {http://dx.doi.org/10.1007/978-3-662-09291-0}
  {\path{doi:10.1007/978-3-662-09291-0}}.

\bibitem{Sharma:2012}
A.~Sharma, S.~Dey, Performance investigation of feature selection methods and
  sentiment lexicons for sentiment analysis, IJCA 3 (2012) 15--20.

\end{thebibliography}

\end{document}